\documentclass{article}
\usepackage[utf8]{inputenc}

\usepackage{url}

\usepackage{graphicx}
\graphicspath{ {images/} }

\usepackage{amsmath}
\usepackage{nicefrac}

\usepackage{nameref}

\usepackage{verbatim}
\usepackage{subcaption}
\usepackage{xcolor}
\usepackage{tocloft}
\usepackage{hyperref}
\hypersetup{
    colorlinks=true,
    linkcolor=blue,
    filecolor=magenta,      
    urlcolor=cyan,
    pdftitle={Overleaf Example},
    pdfpagemode=FullScreen,
    }
\usepackage[toc]{appendix}
\newcommand{\appendixsubsection}[1]{
    \stepcounter{subsection}
    \subsection*{\Alph{section}.\arabic{subsection}\hspace{1em}{#1}}
}

\usepackage{sepfootnotes}
\usepackage{authblk}

\usepackage{amsmath,amssymb}

\usepackage{listings}
\usepackage{xcolor}
\definecolor{codegreen}{rgb}{0,0.6,0}
\definecolor{codegray}{rgb}{0.5,0.5,0.5}
\definecolor{codepurple}{rgb}{0.58,0,0.82}
\definecolor{backcolour}{rgb}{0.95,0.95,0.92}
\lstdefinestyle{mystyle}{
    backgroundcolor=\color{backcolour},   
    commentstyle=\color{codegreen},
    keywordstyle=\color{magenta},
    numberstyle=\tiny\color{codegray},
    stringstyle=\color{codepurple},
    basicstyle=\ttfamily\footnotesize,
    breakatwhitespace=false,         
    breaklines=true,                 
    captionpos=b,                    
    keepspaces=true,                 
    numbers=left,                    
    numbersep=5pt,                  
    showspaces=false,                
    showstringspaces=false,
    showtabs=false,                  
    tabsize=2
}
\usepackage{titling}
\makeatletter
\renewcommand\thefootnote{\textsuperscript{\@fnsymbol\c@footnote}}
\let\old@thanks\thanks % save the normal thanks
\DeclareRobustCommand\thanks[2][]{% redefine thanks
  \AddToHook{begindocument/end}{% but postpone the change until it is needed
    \if\relax#1\relax%
      \footnotemark%
    \else%
      \protect\refstepcounter{footnote}\protect\label{#1}%
    \fi%
    \protected@xdef\@thanks{%
      \@thanks\protect\footnotetext[\the\c@footnote]{#2}%
    }%
  }%
}
\AddToHook{begindocument/begin}{\let\thanks\old@thanks} % present hyperref with the usual thanks
\let\old@maketitle\maketitle
\def\maketitle{\old@maketitle\def\thefootnote{\@arabic\c@footnote}}
\makeatother

\newcommand\independent{\protect\mathpalette{\protect\independenT}{\perp}}
\def\independenT#1#2{\mathrel{\rlap{$#1#2$}\mkern2mu{#1#2}}}

\title{\texttt{balance} - a Python package for balancing biased data samples}
\author{Tal Sarig\ref{equal contributor}}
\author{Tal Galili\ref{equal contributor}} 
\author{Roee Eilat}
\affil{Meta}

\date{June 2023}

\thanks[equal contributor]{The first two authors contributed equally to this work}

\pretitle{\begin{center}\LARGE\includegraphics[width=\textwidth]{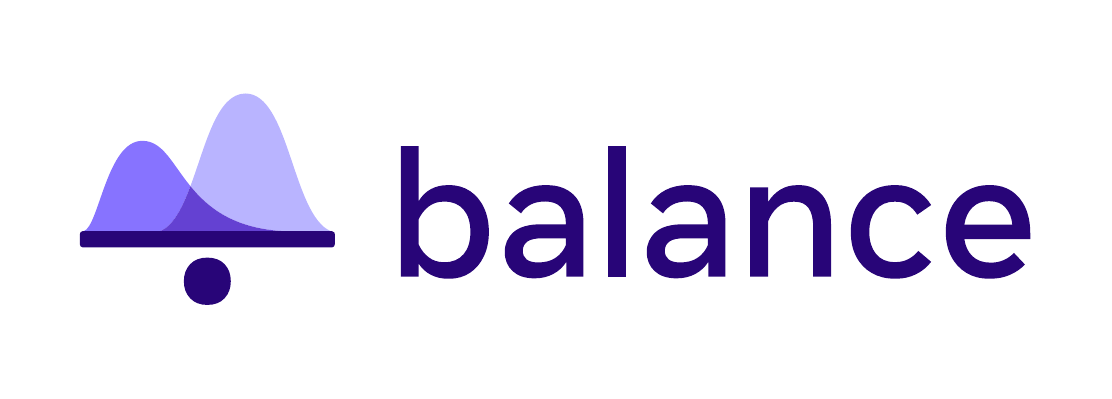}
}
\posttitle{\end{center}}

\definecolor{col_unadjusted}{RGB}{222, 45, 38}       % Red
\definecolor{col_adjusted}{RGB}{52, 165, 48}        % Green
\definecolor{col_target}{RGB}{158, 202, 225}        % Blue

\begin{document}

\maketitle
\vspace{-1.5em}
\begin{abstract}

Surveys are an important research tool, providing unique measurements on subjective experiences such as sentiment and opinions that cannot be measured by other means. However, because survey data is collected from a self-selected group of participants, directly inferring insights from it to a population of interest, or training ML models on such data, can lead to erroneous estimates or under-performing models. In this paper we present \textbf{\texttt{balance}}, an open-source Python package by Meta, offering a simple workflow for analyzing and adjusting biased data samples with respect to a population of interest. 

The \textbf{\texttt{balance}} workflow includes three steps: \textbf{understanding} the initial bias in the data relative to a target we would like to infer, \textbf{adjusting} the data to correct for the bias by producing weights for each unit in the sample based on propensity scores, and \textbf{evaluating} the final biases and the variance inflation after applying the fitted weights. The package provides a simple API that can be used by researchers and data scientists from a wide range of fields on a variety of data. The paper provides the relevant context, methodological background, and presents the package's API.
\end{abstract}

\setcounter{tocdepth}{3}
\tableofcontents

\newpage

\section{Introduction}

% - Surveys: past, present and future.
Surveys play an important role in the study of social phenomena across research fields and industries. From their traditional usage by statistics bureaus in producing population estimates, through the long history of public opinion surveys in political science, to more recent applications like studying user experience in online services and even playing part in epidemiological studies \cite{astley2021global}. The widespread use of surveys, and their unique role in providing measurements on subjective indicators such as sentiment and opinions, makes the field abundant with methodological research.

%- The importance of bias correction in survey based estimation. (non response bias)
A central challenge in designing and analyzing survey data stems from bias due to sampling limitations and non-response. Since the data is collected from a self-selected group of participants, directly inferring insights or training ML models on such data can result in erroneous estimates or under-performing models. An insightful theoretical framework for the sources of error present in survey data is given in the "Total Survey Error" framework \cite{groves2010total}. While the sources might be different, similar manifestations of bias are often present in observational studies when comparing treatment groups, and in any data produced through self-selection processes.

% - Notable methodologies in high dimensional analysis (Raking, IPSW, Multilevel regression with poststratification (MRP), CBPS).
The field of survey statistics offers methods for mitigating bias in samples, at least partially, by relying on auxiliary information (i.e., “covariates” or “features”). When such information is available for all items in the sample as well as for the population from which it was sampled, it can be used to create weights. Under some assumptions on the relation between the auxiliary information, the response mechanism, and the survey responses, applying the weights to the data will produce less biased estimates or models. Different approaches were proposed for the task, from simple post-stratification \cite{little1993post} to methods more suitable for high dimensional covariates space such as raking \cite{deville1992calibration, kalton1983compensating, mercer2018weighting}, inverse propensity weighting \cite{david1983nonrandom, little1986survey, ekholm1991weighting}, covariate balancing methods \cite{imai2014covariate}, outcome regression based approaches \cite{gelman2006data}, and others. Weighting methods have been shown to be effective in reducing bias of survey estimates \cite{solon2015we}.

% - Python as an increasingly prevalent programming language for data science and the need for software tools.
Following methodological advancements in survey statistics, statistical software packages were developed to allow researchers and practitioners to apply these methodologies to survey data and observational data. Most software packages for this aim have R implementations, and other implementations in environments such as SPSS, stata, SAS exist as well. In recent years a rich ecosystem of data science software has been developed for Python, and its usage has become prevalent among researchers and data scientists. This shift created a need for a reliable Python package for working with survey data, and more generally with biased data sets. Here we introduce \texttt{balance} - a Python package for balancing biased data samples. \texttt{balance} offers a simple easy-to-use framework for weighting data and evaluating its biases. The package is designed to provide best practices for weights fitting and offers several modeling approaches. The methodology in \texttt{balance} can support ongoing automated survey data processing, as well as ad-hoc analyses of survey data.

% Workflow and methodologies The methodologies currently built into the package.
The main workflow API of \texttt{balance} includes three steps: (1) understanding the initial bias in the data relative to a target population as observed by the differences in covariate distribution (2) adjusting the data to correct for the bias by producing weights for each unit in the sample based on propensity scores, and (3) evaluating the final biases and the variance inflation after applying the fitted weights. The adjustment step provides a few alternatives for the researcher to choose from: Inverse propensity weighting using logistic regression model based on LASSO (Least Absolute Shrinkage and Selection Operator \cite{robert1994regression}), Covariate Balancing Propensity Scores \cite{imai2014covariate}, Raking, and post-stratification. The focus is on providing a simple to use API, based on Pandas's DataFrame structure, which can be used by researchers from a wide spectrum of fields.

% - Our audience
In this paper we describe the \texttt{balance} workflow in more detail and provide guidance on how to implement it using the package. We include details on methods, assumptions, and model choices made in the package. The methodological background part of the paper is an accessible review of the theoretical frameworks, methods, and practices often used in survey statistics. We invite readers new to the field to use it as a short and effective introduction.

% - Structure of the paper.
The rest of this paper is structured as follows. We discuss related work in Section \ref{sec:Related_Work}, focusing on software packages available in the R and Python ecosystems for survey data analysis and related use cases. In Section \ref{sec:methodological_framework} we provide details on the statistical background that guided the implementation of the package, including theoretical frameworks, estimation methods, and diagnostic tools. In Section \ref{sec:balance_workflow} we present the \texttt{balance} workflow and provide an end-to-end walk through using code snippets that are applied to simulated data. We conclude with a discussion on future directions for the package in Section \ref{sec:future_direction}.

\section{Related Work}
\label{sec:Related_Work}

The open-source ecosystem offers a variety of packages for weighting biased data. This section gives a brief survey of prominent tools in this space and describes some of their capabilities. We find the R ecosystem to be the most developed in terms of packages for survey analysis. The Python ecosystem has some packages for survey statistics. It also has several, well developed, packages for casual inference, which employs similar models (e.g.: propensity scores models, outcome models, etc.). While various R packages exist with similar capabilities to what is available in \texttt{balance}, no Python package (that we are aware of) gives a comprehensive end-to-end coherent solution for researchers.

% \subsubsubsection{\textbf{Survey statistics packages in R}}

The R ecosystem is exceptionally rich and diverse when it comes to survey statistics. The most comprehensive review can be seen in the CRAN task view of "Official Statistics \& Survey Statistics" \cite{cran_taskview_official_stats}. To date, it includes over 130 packages - ranging from the classical \texttt{survey} package \cite{survey_r_pkg_paper} to more niche packages. Similarly, the CRAN task view of "Causal Inference" \cite{cran_taskview_causal_inference} also includes over 130 packages that offer related methods. A short review on the current state of R packages can be found in the \texttt{PSweight} R package \cite{PSweight_paper}, which compares 9 R packages that implement propensity score weighting with discrete treatments. For survey weights diagnostics, the \texttt{cobalt} package \cite{greifer2020covariate} offers many options, including balance tables and plots for covariates of multiple groups. This package includes various capabilities that could inspire future development of \texttt{balance}.

% \subsubsubsection{\textbf{Survey statistics packages in Python}}

For Python, the ipfn package \cite{ipfn} specializes in implementing a fast iterative proportional fitting (raking). This package is utilized in \texttt{balance} and used as the back-end for the raking implementation we rely on. The \texttt{quantipy3} package \cite{Freysson_Quantipy3} is designed to support data processing, analysis and reporting for survey data using pandas and numpy. It supports native handling of special data types like multiple choice variables, statistical analysis using case or observation weights, DataFrame metadata and different data exports. \texttt{quantipy3} seems to be the most similar to what \texttt{balance} tries to achieve but lacks many of the capabilities \texttt{balance} has in all stages of the workflow. The \texttt{samplics} package offers a comprehensive solutions for dealing with complex sampling designs \cite{diallo2021samplics}, with various overlapping and non-overlapping capabilities between this package and \texttt{balance}. The package offers tooling for random selection techniques used to draw a sample from a population, and sample size calculations. It also provides methods for weight adjustments using post stratification or calibration. Additional capabilities in \texttt{samplics} include functions for estimation of statistics and their variance (beyond just the Taylor linearization estimation in \texttt{balance}). These include bootstrap, balanced repeated replication and Jackknife. Other packages we found seem to be only lightly maintained, and do not provide additional capabilities of relevance to our use-case. These include PySurvey \cite{Friedman_PySurvey}, Surveyweights \cite{Wildeford_Surveyweights}, pscore\_match \cite{pscore_match}, pymatch \cite{pymatch}, and causal\_nets \cite{causal_nets}.

% \subsubsubsection{\textbf{Python casual inference packages}}

Stepping aside from survey statistics, several Python packages offer tools for casual inference that can be repurposed for adjusting biased samples.
The \texttt{DoWhy} package \cite{dowhypaper}, developed by  Microsoft, is a well maintained package with the focus on causal inference. It models a given problem as a causal graph to help explore assumptions clearly, estimate causal effects, and test assumptions' validity for robustness. It offers a variety of methods for estimation including Propensity-based Stratification, Propensity Score Matching, and Inverse Propensity Weighting (similar to \texttt{balance}). It also offers outcome based models (currently not implemented in \texttt{balance}) using Linear Regression or Generalized Linear Models, and supports other methods such as Instrumental Variable methods. The package emphasizes graphical interpretation of causal inference. It also gives various refutation methods (dummy outcome, simulated outcome, etc.) and basic visualizations (e.g.: barplots of treatment and control). The \texttt{Empirical Calibration} package \cite{wang2019empirical}, developed by Google, provides a method to compute empirical calibration weights using convex optimization. This approach balances out the marginal distribution of covariates directly while reducing the inflation of variance. This is similar to performing raking while trying to keep the weights to be as equal as possible. It offers a bias correction solution that resembles the raking and CBPS methods that are implemented in the balance package. The \texttt{causalml} package \cite{chen2020causalml} provides a set of modeling and causal inference methods for analyzing observational data using machine learning algorithms. It provides tool to estimate the Conditional Average Treatment Effect (CATE) and the Individual Treatment Effect (ITE). This package offers a wide variety of ML algorithms, including tree-based algorithms, meta-learner algorithms, instrumental variables algorithms, and neural-network-based algorithms. While these packages are comprehensive, there is still an overhead and complexity for using them for balancing data for the workflow \texttt{balance} is optimized in handling with the focus on surveys data.

\section{Methodological Background}
\label{sec:methodological_framework}

Before diving into the workflow and the implementation details of \textit{balance}, we introduce a brief description of the methodological background concerning the representation error problem in surveys, weights estimation and tools to evaluate survey weights.

\subsection{The Total Survey Error framework}
\label{subsec:total_survey_error}

\begin{figure}[ht]
    \centering
    \includegraphics[width=\textwidth]{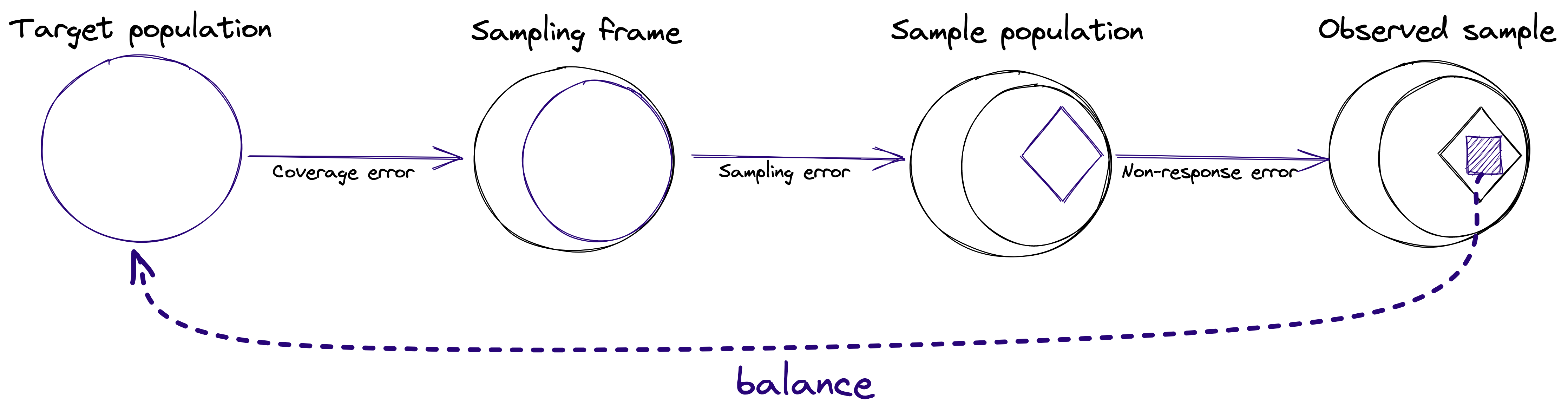}
    \caption{A flow diagram of "Total Survey Error", illustrating the different components of surveys' representation error.}
    \label{fig:total_survey_error_flow}
\end{figure}

The "Total Survey" Error framework \cite{groves2010total} provides a theoretical framework to describe statistical properties of surveys. It is used as a conceptual tool for researchers when designing and analyzing surveys to minimize estimation errors and biases. While the research goal is to estimate a population parameter, such as average or ratio, surveys only provides a glimpse on this parameter through the survey responses and are subject to a range of sources for statistical errors, as described by the "Total Survey Error" concept. 

The "Total Survey Error" has two main components: representation error and measurement error \cite{salganik2019bit}. Since neither can be overcome by increasing the sample size, researchers should be aware of these as early as the survey design stage. \textit{Measurement error} deals with potential biases introduces to the estimation due to the instrument of measurement. It includes questions about the validity of the responses, the phrasing of the questions and how it is affecting what we are trying to estimate, and similar questions related to whether we measure the exact quantity we aim for. \texttt{balance} is focused on addressing and correcting the representation errors in this framework, and hence for the rest of the section we will focus on the representation part of the framework. 

The \textit{representation error} deals with how to infer from a subset of people to the whole population on which we would like to learn, referred to as the \textit{target population}. The magnitude of the error depends on the group of respondents to the survey and depends on how similar or different this group is from the target population. Formally, Figure \ref{fig:total_survey_error_flow} shows the different sources of representation error and illustrates a breakdown of the difference between the group of respondents and the target population.

The first error we consider is the \textit{coverage error}. Its driver is the misalignment between who can be sampled for the survey (the "sampling frame") and the target population. In today's world, where many, if not most, surveys are conducted through the internet, a common sampling frame is people with access to the internet. Since this sampling frame may not be representative of the whole population, a caution should be taken when conducting survey over the web.   

The most canonical example for a sampling frame that is not fully covering the target population is the "Literary Digest" 1936 poll \cite{freedman2007statistics}. During this year the Literary Digest magazine ran a poll to predict the result of the U.S. election. Franklin Delano Roosevelt was the Democrats candidate and the Republican candidate was Governor Alfred Landon of Kansas. The magazine predicted a decisive victory for Landon with a poll that was based on roughly 2.4 million voters, but, as history tells us, Roosevelt won 62\% of the votes. Even though the poll sampled 10 million people, the sampling frame was skewed. The sample included magazine readers, and people from phone lists and club memberships lists. However, since people of lower socioeconomic status were disproportionately not part of the magazine's audience those days, the poll missed a significant and unique portion of the U.S. voters population Due to setting a sampling frame that ignores the target population definition, the magazine's coverage error was large and led to mis-prediction of the elections' results.

Once the sampling frame is set, the researcher samples a certain amount of people from the frame to ask to reply on the survey. This is the \textit{sample population}, or the group of people that have a "real" opportunity to reply on the survey. When doing so, the researcher reveal another gap where error can occur due to sampling, which is the \textit{sampling error}. This might be small if we are able to sample either completely at random from the sampling frame or by designing the sampling with the right sampling probabilities, but can be significant given wrong assumptions on the structure of the sampling frame or a complex mechanism of sampling. This error can be reduced as we increase the sample size (and will be 0 if the sample population is the same is the sampling frame), and is the one captured by the margin of error, often reported with a survey results.

Once researcher has sent out the invitation for filling the survey to the sample population, most often only a portion will choose to take part in the survey. These are the respondents, or the \textit{observed sample}. This self selection behavior causes another error component which is the \textit{non-response error}. This error can be substantial depending on how the survey is conducted, the survey questions, and other issues related to the instrument. The percent of non-response can give us some intuition of how large this error is but the actual size of the bias depends only on the properties of the people who chose to respond. In the case of the Literary Digest poll the response rate was only 24\%. In fact, research suggests that the primary source of the error in the poll originated in the non-response bias. Specifically, people who strongly disliked Roosevelt were more willing to take the time to mail back their response \cite{squire1988why, lusinchi2016president}.

\texttt{balance} aims to correct for all types of representation errors at once (see Figure \ref{fig:total_survey_error_flow}). Using additional assumptions, as described in the next section, we are able to make the group of respondents, i.e. the observed sample, similar in properties to the target population and hence overcome some parts of the representation error. However, it is important to note that there are cases where it is impossible to fully correct the representation error. Such cases occur when the assumptions on the missingness are not satisfied. The simplest example of such case is when there is a substantial coverage error for which we cannot overcome using auxiliary data. For example, if we want to learn about North America's population but survey people only from The United States. Even given lots of auxiliary information we will likely not be able to adjust the sample such that it correctly represents Canada's population as well.

\subsection{Definitions and notations}

With the Total Survey Error framework in mind, we will now set definitions to be used throughout the paper. Let $\mathcal{S}$ denote a sample of respondents to a survey consisting of $n$ respondents (sometimes referred to as the sample), and let $\mathcal{T}$ represent a target population with $N$ units.

Furthermore, we assume we have some auxiliary data on all units in sample and target population, represented by a covariates (or features) vector attached, $X_i$. Note that we assume that the same covariates are available for the sample and the target, otherwise we ignore the non-overlapping covariates. This framework is applicable when we have the auxiliary data at the unit level for the entire population or for a representative random sample from it. For example, when we sample from a list of customers for which we have auxiliary information available, or in cases when a reference survey is available (a survey with better sampling properties to be used for correcting biases \cite{lee2009estimation}). Another common use-case is when census data of the population is available.

We define $R$ to be an indicator for inclusion in the respondents group\footnote{For estimation in \texttt{balance}, we think of the target population as a reference group, and hence units of the target population are distinct from the units of the observed sample.}, i.e. $R_i=1$ if $i \in \mathcal{S}$ and $R_i=0$ if $i \notin \mathcal{S}$. Furthermore, we define $y$ to be the answer to one item of the survey. The answer can be numeric or discrete, and is observed only for $\mathcal{S}$. In our setup, we think about $y$ as a constant and the random variable, later considered for statistical properties, is $R$.

Our objective is to estimate a certain parameter of the population. The simplest example is estimating the mean of one item of the survey, i.e. the mean of $y$. In this case, a natural estimate to $\bar{y}=\frac{1}{N}\sum_{i\in\mathcal{T}}y_i$ is the sample mean $\bar{y}_\mathcal{S}=\frac{1}{n}\sum_{i\in\mathcal{S}}y_i$. However, due to the non-random sampling of $\mathcal{S}$ from the population $\mathcal{T}$, the proposed estimate will be biased, such that $\mathbb{E}\left[\bar{y}-\bar{y}_\mathcal{S}\right] \neq 0 $.

\subsection{Estimation of the survey weights}

Weights are a common way to overcome survey error, and are essential when estimating a population parameter \cite{solon2015we}. This is generally done by incorporating the weights, $w_i$ where $i \in \mathcal{S}$, into the parameter estimation procedure\footnote{ In the \texttt{balance} package, we choose to scale the weights to sum to the population size. This way, each sample unit weight represents the number of corresponding units from the population.}. An example is the case of estimating a parameter of the population using a weighted mean of the sample: 
\begin{equation}
\bar{y}_w = \frac{\sum_{i=1}^{n} w_i y_i }{\sum_{i=1}^{n} w_i}
\end{equation}
Further details about using the weights for estimations are described in section \ref{subsec:diagnostics_outcome}.
One of the advantages of using weights, unlike alternative methods, is the flexibility it gives in estimation. The weights depend only on the group of respondents and not on the outcome itself, hence give the researcher the flexibility to use the same set of weights for multiple outcomes, combine multiple outcomes into one parameter, or consider the outcome in only in specific cuts of the population.

We typically employ weights to adjust from the observed sample to match the target population, so to overcome the representation bias of the sample. In scenarios where the sampling procedure is set by design and therefore known, we define the inverse of the sampling (or selection) probabilities as \textit{design weights}\footnote{We may assume each unit $i$ in $\mathcal{S}$ and/or $\mathcal{T}$ has a design weight, $d_i$. These are the sampling weights that are based on the known sampling procedure from the sampling frame to the sample population. These are often set to 1 for the respondents when the sampling probabilities are unknown.}. However, to overcome the full gap between the respondents and the target population we need to estimate the weights according to the actual realization of the observed sample. When estimated against the complete target population, the weights can help address the non-response error, the by-design sampling error, the "unknown" sampling error and the coverage error.

%% TODO: <start moving text from here
% TODO: I'm proposing to just move (and merge) this section with the section about "Diagnostics for the outcome". And leave here only a sentence saying that the weights can be used for weighted mean and variance, and that more details are in the "Diagnostics for the outcome" section (and give ref). TalS/Roee - WDYT? I moved part of this paragraph to the begiinng and references to the relevant section.
%For some set of weights $w_i$, corresponding to the inverse of the sampling probability of each unit of the sample $i \in \mathcal{S}$, and a set of survey responses $y_i$, the estimated weighted mean of the parameter of interest in the population, using a Horvitz–Thompson estimator \cite{horvitz1952generalization}, is given by:

%\begin{equation}
%\bar{y}_w = \frac{\sum_{i=1}^{n} w_i y_i %}{\sum_{i=1}^{n} w_i}
%\end{equation}

%The corresponding variance estimation of the estimated weighted mean, when weights are assumed to be known (i.e.: fixed), is then:

%\begin{equation}
%\widehat {V (\bar y_w)} 
% = \frac{1}{(\sum_{i=1}^n w_i)^2} \sum_{i=1}^n %w_i^2 (y_i - \bar y_w)^2
%\end{equation}

%The derivation of this estimator can be found section \ref{subsec:var_of_weighted_mean} (\nameref{subsec:var_of_weighted_mean}).
%% TODO: until here is the part of the text I think should be moved/merged>

A few assumptions are required to utilize the estimated weights for valid estimations of the parameter of interest and mitigate the representation bias. Under these assumptions, the estimation of the weights relies on the auxiliary data $X_i$.

The first assumption is the Missing At Random assumption (MAR) \cite{rubin1976inference}. The MAR assumption states that the response mechanism is independent of the survey responses conditional on the auxiliary data. In other words, $Y \independent R \mid X$, which means that given the covariates the likelihood of a person to respond to the survey doesn't depend on their answer. This assumption is also known as the ignorability assumption or conditional unconfoundedness in causal inference literature \cite{rosenbaum1983central}. It is worth noting that recent research (such as \cite{little2021missing}) proposes alternative approaches to address missing values created by design in surveys analysis. 

The second assumption is positivity $0<P(R_i=1|X_i)<1$ for all units in $\mathcal{S}$ and $\mathcal{T}$. $0<P(R_i=1|X_i)$  means that given the auxiliary data, every unit of the target population has non-zero probability to be include in the sample. In other words, in a counterfactual world, any unit $i \in \mathcal{T}$ could have participated in the survey given their covariates. Conversely, we also assume $P(R_i=1|X_i)<1$, implying that every unit in the observed sample is also in the target population. The combination of the MAR assumption and the positivity assumption is often known as the "strong ignorability" assumption \cite{rosenbaum1983central}.

Given the assumptions we are now left with the question of how to estimate the weights. The \texttt{balance} package currently supports 4 different methods for estimating the weights: post-stratification, raking, Inverse Propensity score Weights (IPW or IPSW) and Covariate Balancing Propensity Score (CBPS). Next, we provide more background about the estimation process in each method and describe advantages and limitations of each.

\subsubsection{Post-stratification}

Post-stratification \cite{little1993post} is one of the most common weighting approaches in survey statistics. It originates from a stratified sample or probability sampling, where the population is divided into sub-populations (strata) and a sample is independently drawn from each. However, in post-stratification, the stratification is done after the sample has been selected. This is done to overcome errors originating in mechanisms outside the sampling design, such as non-response.

The idea behind post-stratification is straightforward. For each cell (strata) in the population, calculate the percentage it represents of the total population. Then fit weights so that they adjust each stratum in the sample so to have the same proportions as each strata as in the population.

Let $\mathcal{H}$ be the group items which represent some stratum in the population, and $P_\mathcal{H}$ represent the proportion of this stratum in the target population $\mathcal{T}$, i.e. $p_h=\frac{|\mathcal{H}|}{N}$. Let $n_\mathcal{H}$ be the number of respondents from stratum $\mathcal{H}$  in the observed sample. We also define the "inflation factor" as $I = N/n$, i.e. the factor indicating by how much we need to multiply the total observed sample size to get to the total target population size. Consequently, the post-stratification weight for each unit $i$ from stratum $\mathcal{H}$ in the observed sample is:

\begin{equation}
w_i = P_\mathcal{H}\frac{n}{n_\mathcal{H}} * I \quad \forall i \in \mathcal{H}
\end{equation}

Note that the multiplication by $I$ is a result of the arbitrary choice to scale the weights to the population size, and could be omitted.

The goal of post-stratification is to have the sample exactly match the joint-distribution of the auxiliary data of the target population. Hence it requires the researcher to know the joint distribution of the covariates to weight on. This level of resolution for the target population may not always be available. When only marginal distributions of the covariates are available then raking might serve as an alternative method to estimate the weights. Raking is described in sub-section \ref{subsec:raking}.

Another limitation of post-stratification is on the number of covariates that can be used to correct the biases due to the limitations of having enough respondents in each of the cells. Having a cell with very few respondents could easily lead to a handful of respondents that receive very large weights - which leads to inflated variance of estimation based on such weights. Furthermore, when continuous variables are required for weighting, the researcher must decide on the thresholds for bucketing the variables into. A more general approach is the inverse propensity score weighting described in sub-section \ref{subsec:inverse_propensity_core_weighting}.

\subsubsection{Raking}
\label{subsec:raking}

\textit{Raking} \cite{deville1992calibration, kalton1983compensating, mercer2018weighting}, also known as Iterative Proportional Fitting procedure (IPF), is a method that fits the sample data to a target population using only the marginal distributions of the population's covariates. Typically, we have access to these marginal distributions but often not to their joint distribution. Since raking weights do not represent the joint distribution, this can be thought of as a type of regularized model. This approach helps to avoid over-fitting small cells as in post-stratification and instead focuses only on the marginals \cite{battaglia2009practical}. 

Raking essentially applies post-stratification sequentially over all covariates using only the marginal distributions. This is done repeatedly until a convergence is achieved. If exist, the design weights of the sample are used as the starting point of the algorithm. For example, we may have the marginal distribution of gender, age, and education. Raking would first adjust weights to match the gender distribution and then take these weights as input to adjust for age, and then for education. It would then adjust again to gender and then again to age, and so forth until it converges. This process will repeat until one of three stopping criteria are met: (1) we reached a pre-defined number of iterations,(2) the maximum difference in proportions between sample and target marginal distribution on any covariate is smaller then a preset convergence rate, or (3) the weights have converged and the change from one iteration to another is smaller then a preset rate tolerance parameter.

The resulting weights will be close to the marginal distribution of the population covariates. However, one cannot assume that the weighted sample joint distribution is the same as the joint distribution in the target population. Hence, if one wants to infer only for a sub-group of the population (such as young-adults only), it is less recommended to use raking weights, and, if possible, one should prefer a method that take into account the joint distribution of the covariates if such data exists. 

Similar to post-stratification, raking is limited by the number of covariates that can be included in the model, due to the need to have enough respondents in each margin cell. In addition, raking may be sensitive to the order in which  covariates are adjusted for, which may lead to under-correction of some covariates.

\subsubsection{Inverse Propensity score Weighting (IPW)}
\label{subsec:inverse_propensity_core_weighting}

A natural expansion to post-stratification and raking is the inverse propensity score weighting that can be viewed as a continuous extension of post stratification. 

The \textit{Propensity Score} is defined as the conditional probability to be part of the observed sample given the covariates: 
\begin{equation}
p(X):=Pr(R=1|X)
\end{equation}
It was first suggested by Rosenbaum and Rubin \cite{rubin1976inference, rosenbaum1983central}  as a method to perform matching for causal effects estimations in observational studies, and was later adopted to weighting survey data \cite{david1983nonrandom, little1986survey, ekholm1991weighting}. Rosenbaum and Rubin \cite{rosenbaum1983central} showed that the assumptions of "strong ignorability" (uncounfoundness: $Y \independent R \mid X$ and positivity: $0<P(R_i=1|X_i)<1$) implies that $Y \independent R \mid p(X)$, and that $p(X)$ is the coarsest balancing score (a score $B(X)$ that satisfies $Y \independent R \mid B(X)$). This means that the propensity score is an inexpensive way, in terms of dimension, to estimate the selection probabilities. Hence, in the spirit of "Horvitz–Thompson estimator" \cite{horvitz1952generalization} of using the inverse selection probabilities as weights, the inverse of the propensity score was suggested as a weighting procedure to adjust for non-response bias \cite{little1986survey}.
 
The estimation of the propensity scores can be done in any standard tools for classification, such as logistic regression, decision trees and random forests (such as in \cite{watkins2013empirical}) or others. The choice of the model depends on the researcher assumptions regarding the parametric model of the non-response and the number and types of features used. In \texttt{balance}, we chose to implement the estimation of the propensity scores through a (regularized) logistic regression. The logistic regression model assumes a linear relation between the covariates and the log odds, of the form: 
\begin{equation}
\log(\frac{p_i}{1-p_i})=\beta^T X
\end{equation}

Once the propensity scores are estimated, the weight of unit $i$ is calculated by  $w_i=\frac{1-\hat{p_i}}{\hat{p_i}}$. This is because we define the target population as a reference group and we don't assume the target doesn't include the observed sample (i.e. we don't exclude units from the target based on their appearance in the sample). In this case, borrowing concepts from the causal inference literature \cite{li2018balancing}, the estimation we care about is only the estimation of the average treatment effect for the "control" ("untreated") group (the target population), and hence we use $\frac{1-\hat{p_i}}{\hat{p_i}}$ as the weights. 
%TODO: add a proof in the appendix?

One challenge when including many covariates is that the estimation of the propensity scores (and hence, the weights), can have a a high variance, which may lead to unnecessary inflation of the survey estimates\footnote{Note that the variance estimator of the weighted mean presented in subsection \ref{subsec:diagnostics_outcome} is a closed form formula that assumes fixed weights, and hence the variability in the estimation of the weights is not reflected in the formula. A more accurate estimator of the variance would rely on bootstrap samples, which are more computationally expensive.} \cite{little2005does}. In \texttt{balance} we try to mitigate this by applying regularization to the logistic model using LASSO (Least Absolute Shrinkage and Selection Operator) \cite{tibshirani1996regression}. This either excludes or reduces the magnitude of the covariates' coefficients that are not predictive for the response mechanism in the propensity model. This helps to minimize the variance of the estimated weights, at the potential expense of some consistent (hopefully small) bias in their estimated values. However, this process doesn't exclude covariates that are uncorrelated with the response itself. These should be excluded by the researcher in order to avoid variance inflation \cite{little2005does}. Another protective measure against variance inflation and extreme weights is weight trimming. \texttt{balance} offers automatic trimming, for details see subsection \ref{subsec:How does balance implement the adjustment?}. 

Another weakness of inverse propensity score weighting is that it may strongly depend on the specification of the model of the propensity scores, as shown in a simulation study in \cite{imai2014covariate}. Imai and Ratkovic \cite{imai2014covariate} have suggested the method of Covariate Balancing Propensity Score (CBPS) (described in the next subsection) to overcome this issue. Fitting tree-based methods for the propensity scores have also shown to be a good alternative \cite{lee2010improving, watkins2013empirical}.

\subsubsection{Covariate Balancing Propensity Score (CBPS)}
Covariate Balancing Propensity Score (CBPS), suggested by Imai and Ratkovic \cite{imai2014covariate}, is a method to estimate the propensity score in a way that will also result in maximizing the covariate balance. The method is preferable in the cases of misspecification of the propensity score model, which may lead to a bias in the estimated weights (and consequently, the estimated survey statistic). CBPS is described in details in \cite{imai2014covariate} and implemented in the R package \cite{CBPS_R_package}. We give here a short summary of the method for completeness of the estimation methods section. 

The CBPS method is an expansion of the maximization problem of logistic regression. The propensity score of the logistic regression model is modeled by: 
\begin{equation}
p _\beta(X_i)=\frac{\exp(\beta ^T X_i)}{1+\exp(\beta ^T X_i)} \quad \forall i \in \mathcal{S}, \mathcal{T}
\end{equation}

By the maximum-likelihood approach, $\beta$ is estimated by maximizing the log-likelihood, which results in:
$$
\hat{\beta}_{MLE}=\arg\max_\beta \sum_{i \in \mathcal{S}} \log(p_\beta(X_i))+\sum_{i \in \mathcal{T}}\log(1-p_\beta(X_i))
$$

At the maximum of the log-likelihood $\beta$ satisfies first order condition:
$$
\frac{1}{n} \left[ \sum_{i \in \mathcal{S}} \frac{p^\prime_\beta(X_i)}{p_\beta(X_i)} + \sum_{i \in \mathcal{T}} \frac{-p^\prime_\beta(X_i)}{1-p_\beta(X_i)} \right] =0
$$
where $p^\prime_\beta(X_i)$ is the derivative of $p$ by $\beta^T$.
This condition can be viewed as the condition that balances a certain function of the covariates, in this case the derivative of the propensity score $p^\prime_\beta(X_i)$.

Generally, one can expand the above to hold for any function $f$ of the covariates $X$ ($f(X)$) and depends on the researcher's goals and assumptions:
\begin{equation}
\mathbb{E} \left\{ \sum_{i \in \mathcal{S}} \frac{f(X_i)}{p_\beta(X_i)} + \sum_{i \in \mathcal{T}} \frac{f(X_i)}{1-p_\beta(X_i)} \right\} =0
\end{equation}
CBPS method chooses $f(X)=X$ as the balancing function $f$ in order to balance the first moment of each covariate in addition to the to derivative of the propensity score parametric model. The estimation of the propensity score is then done by using Generalized Methods of Moments (GMM) \cite{hansen1982large} and is described in \cite{imai2014covariate}.

\subsection{Evaluation of survey weights}
\label{sec:diagnostics}

\subsubsection{Overview}
\label{subsec:diagnostics_overview}

As mentioned, survey weights are essential to improve the accuracy of survey estimates, but their reliability and validity hinge on several assumptions and modeling decisions.

Survey weights are valuable when: (a) the non-response pattern is sufficiently captured by the measurable covariates, (b) the covariates are accurately represented in the fitted propensity score model, ensuring that the weighted distribution of covariates in the sample closely resembles that in the target population, and (c) the survey weights correlate with the outcome of interest to an extent that justifies the increased variance resulting from the weights \cite{kish1992, little2005does}.

To see the level to which survey data empirically complies with the above criteria, diagnostics measures can be applied on each of the main elements: covariates, weights, and outcome.

Both covariates and outcomes can be checked before and after applying the weights, allowing for a comprehensive assessment of the weights influence on the data. Such evaluations helps confirm whether the weights have successfully enhanced the representativeness of the sample data in a way that also substantially influences the outcome of interest. Additionally, various diagnostics can be performed on the weights themselves to understand if their are extreme weights which dominate the sample, as well as the overall impact the weights have on the effective sample size.

Distributions can be compared using summary statistics and plots, with calculations incorporating the fitted survey weights. The following sections describes various methods for that purpose.

\subsubsection{Visualizing Distributions}
\label{subsec:diagnostics_visualizations}

\begin{figure}[htbp]
  \begin{subfigure}[t]{0.85\textwidth}
    \centering
    \includegraphics[width=\linewidth]{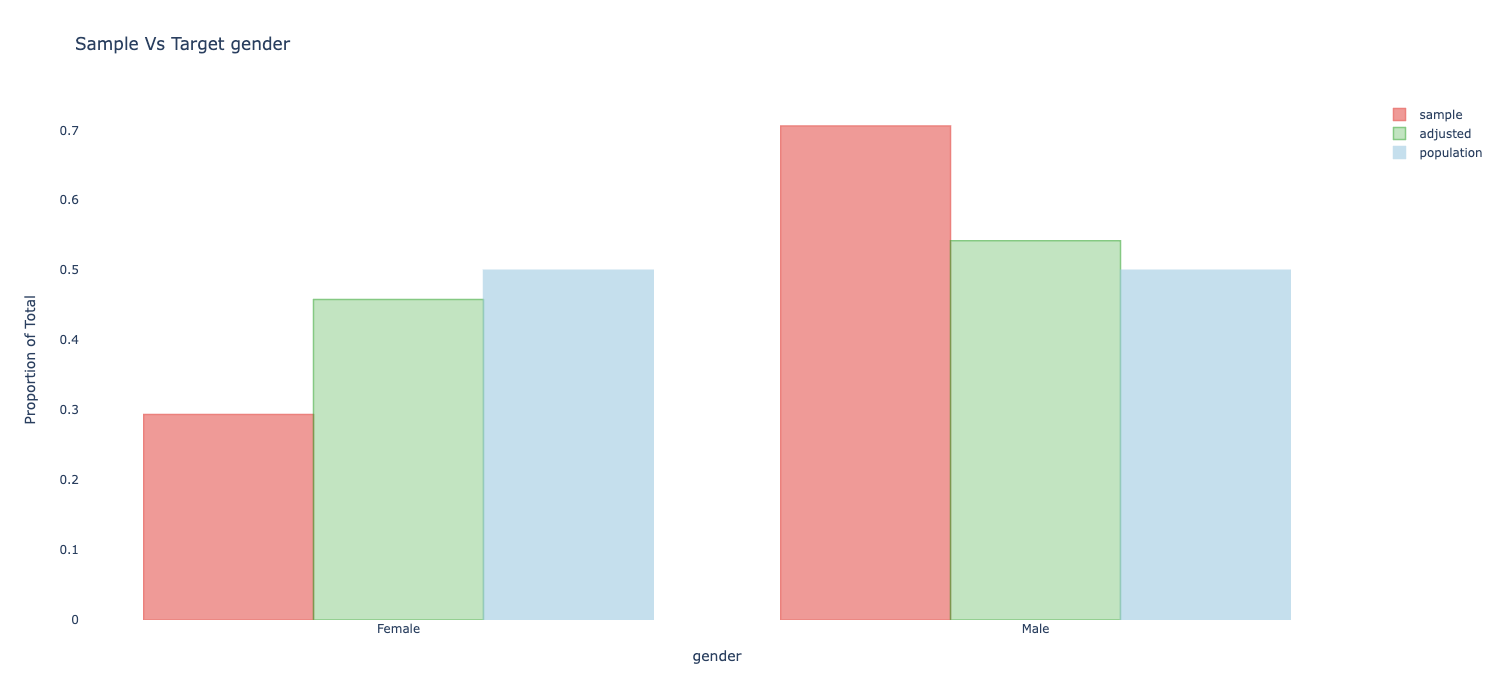}
    \caption{Bar plot for gender}
  \end{subfigure}
  \hfill
  \begin{subfigure}[t]{0.85\textwidth}
    \centering
    \includegraphics[width=\linewidth]{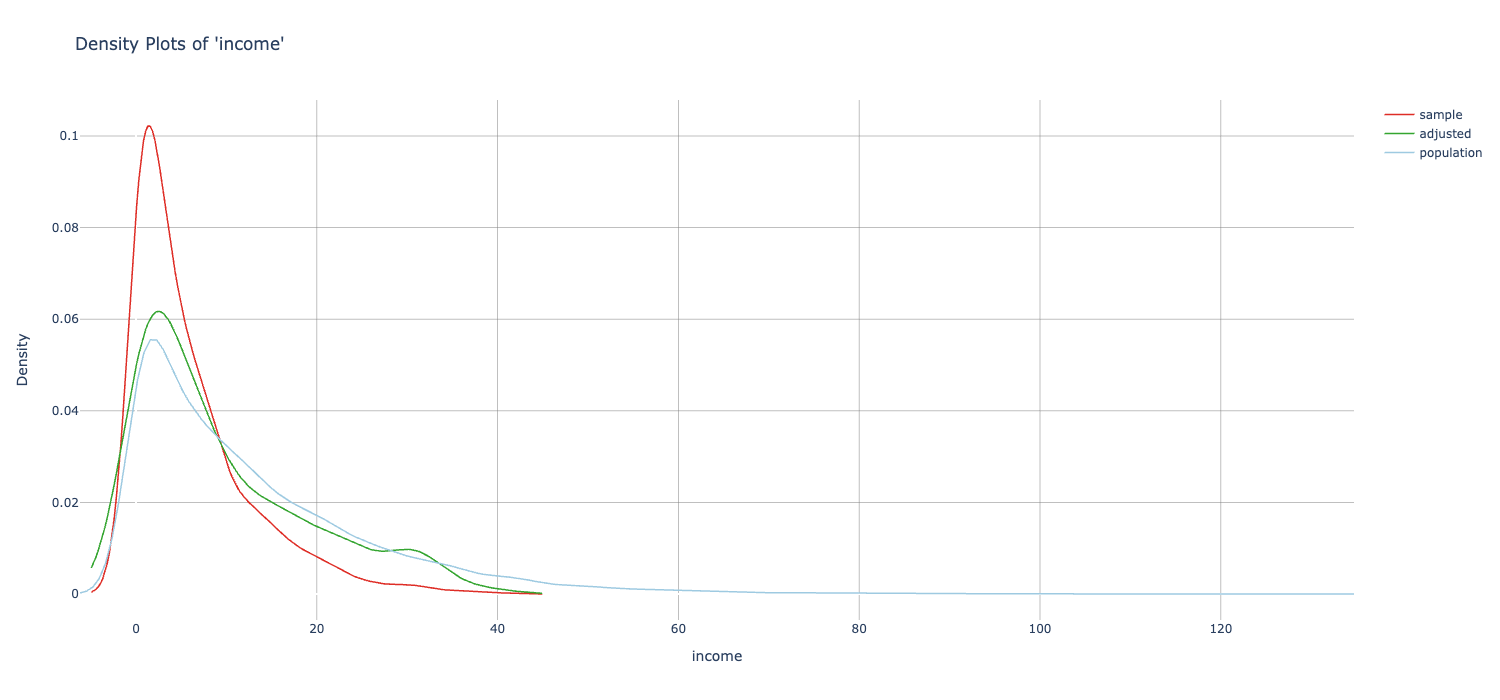}
    \caption{KDE for income}
  \end{subfigure}
  
  \vspace{5mm}
  
\centering
\scriptsize % Adjust the font size here
\begin{tabular}{cccc}
    & \textcolor{col_unadjusted}{\rule[0.5ex]{1em}{1em}} & \textcolor{col_adjusted}{\rule[0.5ex]{1em}{1em}} & \textcolor{col_target}{\rule[0.5ex]{1em}{1em}} \\
    & Un-weighted sample & Weighted sample & True value \\
\end{tabular}
\caption{Examples (from simulated data) of diagnostic plots for covariates}
\label{fig:example_of_diag_plots}
\end{figure}

Distribution plots are effective tools for visualizing the covariates and outcomes in the data, offering insights that extend beyond basic summary statistics. For numerical variables there are Kernel Density Estimator (KDE) plots  (see an example in Fig \ref{fig:example_of_diag_plots}), histograms, and quantile-quantile (QQ) plots. For categorical variables it is common to use bar-plots (see an example in Fig \ref{fig:example_of_diag_plots}). These distribution plots enable users to observe the differences between the observed sample and the target population, as well as the influence of the applied weights.

The advantage of visualizations lies in their ability to reveal unexpected patterns in the complete range of data, as opposed to looking on summary statistics only. However, scaling these visualizations can be challenging. For instance, while examining KDE plots for each covariate comparing the sample and target population is informative, it is often more efficient for the researcher to have summary statistics that can quickly convey the extent of bias in different features. This is particularly useful when evaluating multiple weighting solutions. The following sections discuss particular summary statistics that helps in addressing this need.

\subsubsection{Diagnostics for the covariates using ASMD}
\label{subsubsec:diagnostics_covars_asmd}

A fundamental statistic for comparing distributions is the first moment, i.e. the mean, of each covariate for the target and the sample (weighted or unweighted). For each covariate, it is insightful to observe how much closer the application of weights brings us to the target mean. The \textbf{Absolute Standardized Mean Deviation (ASMD)} can be used to summarize this effect.

The Absolute Standardized Mean Deviation (ASMD) is a statistical measure employed to compare the means of two groups (in our case, the sample and the target). It is computed as follows:

\begin{equation} 
ASMD = \frac{\left| \bar{X}_{Sample} - \bar{X}_{Target} \right| }{SD} 
\end{equation}

where $\bar{X}_{Sample}$ and $\bar{X}_{Target}$ are the means of the sample and target. The $SD$ can be either the pooled standard deviation of sample and target, or the standard deviation of the target population. In \texttt{balance} we use the standard deviation of the target population.

The concept of ASMD is derived from the standardized mean difference, which is a measure of effect size used to compare the means of two groups, expressed in the standard deviation units. This is often referred to as Cohen’s $d$ \cite{cohen2013statistical}, a standardized measure of the magnitude of the difference between two means. ASMD values range from 0 to infinity, with larger values indicating greater differences between the means of the two groups.\footnote{A value of 0 signifies no difference between the means, while a value of 1, for example, indicates that the difference between the means is equal to one standard deviation. ASMD is most easily conceptualized when the distributions being compared are unimodal and symmetric.}

The ASMD can be calculated using the unweighted and weighted mean of the Sample, and these two quantities can be compared. If applying the weights lead to an ASMD value that is closer to 0 than the ASMD of the unadjusted sample, then it is an indication the weights help to reduce the bias. The level of adjustment can be measured by taking the difference of these two ASMD values.

\begin{equation} 
ASMD_{diff} = ASMD_{unadjusted} - ASMD_{weighted}
\end{equation}

The more the adjusted ASMD (the ASMD that is based on the weighted mean of the sample) is smaller than the unadjused ASMD (based on the unweighted mean) - i.e.: the closer the diff is to 0 - the stronger the indication we have of the potential benefit of the weights for adjusting a bias in the covariates. The magnitude of the difference of the two ASMD values is a measure of the impact of the weights. If $ASMD_{diff}$ is positive then it means the weights have helped reduce the bias, while a negative value indicates that the weights have potentially increased the bias.

Since we often wish to adjust over many covariates, then the ASMD difference from each covariate can be summarized by taking the average ASMD (or $ASMD_{diff}$) over all covariates. This gives a single summary statistic to measure the level of impact the weights had on reducing the bias of the sample in the covariates.

For categorical variables, one possible behavior for ASMD calculation is to use dummy variables and calculates the ASMD for each of them. The ASMD per dummy variable approach could lead to over-weighting categorical variables with many categories when calculating the mean ASMD. A possible solution is to aggregate these ASMD per covariate. I.e.: calculates a single mean ASMD value for each categorical variable, and then the general mean ASMD will give each variable the same weight in the final calculation.

% TODO: reference the example below of how it looks.

For some limitations of using ASMD, see the appendix section  \ref{subsec:limitations_of_asmd}.

\subsubsection{Diagnostics for the weights}
\label{subsec:diagnostics_weights}

\noindent \textbf{Kish's design effect}

One important aspect to consider when using survey weights is the potential increase in variance of some estimate of interest (e.g.: the mean) due to the variability in the weights. This is measured by a quantity known as \textit{design effect}, which is generally defined as the variance of the weighted estimator to the variance expected with simple random sample (SRS) without replacement \cite{ARNAB2017645, henry2015design}. It assesses the potential impact that weights might have on the variance of estimating the weighted mean. 

\textit{Kish's design effect} \cite{Kish1965} is a widely known and commonly used design effect measure for the potential impact that weights might have on the variance of estimating the population mean using the weighted mean. Kish's design effect assumes that there is no correlation between the weights and the outcome variable, also known as "haphazard weights.", its formula is:

\begin{equation}
D_{eff} = \frac{n \sum_{i=1}^n w_i^2}{(\sum_{i=1}^n w_i)^2} = \frac{\frac{1}{n} \sum_{i=1}^n w_i^2}{\left(\frac{1}{n} \sum_{i=1}^n w_i\right)^2}
= \frac{\overline{w^2}}{\overline{w}^2}
\end{equation}

The \emph{effective sample size proportion (ESSP)} indicates what is the effective proportion of sample size we'll keep after applying the weights. It's simply the inverse of $D_{eff}$:

\begin{equation}
ESSP = \frac{1}{D_{eff}}
\end{equation}

% TODO: fine reference for ESS?
The \emph{effective sample size} is a related measure that takes into account both the design effect and the actual sample size. It can be used to approximate the number of independent observations that would yield the same variance as the weighted sample. The effective sample size is calculated as follows (where $n$ is the sample size):

\begin{equation}
n_{eff} = ESS = \frac{n}{D_{eff}}
\end{equation}

The effective sample size provides a useful way to gauge the impact of the weights on the precision of the estimates. A smaller effective sample size indicates that the weights have introduced greater variability in the estimates, potentially requiring a larger actual sample size to achieve the desired precision.

Further details on assumptions and proofs are available in appendix \ref{subsec:kishs_deff_appendix}. 

\noindent \textbf{Summary Statistics for the Distribution of Weights}
\label{subsubsec:diagnostics_weights_summary}

While Kish's design effect can be used to estimate an effective sample size as a summary measure for the impact of using weights, it may also be beneficial to examine the distribution of weights using other summary statistics. For instance, extremely large or small weights could indicate potential issues with the weighting process or the presence of outliers in the data used for estimating the weights. Furthermore, the distribution of weights can help determine whether the weights are concentrated on a small number of observations or more evenly distributed across the sample. These observations are often easier to infer from summary statistics than from distribution plots of the weights. Understanding the distribution of the weights can also help to better understand Kish's design effect (and effective sample size) value, which may indicate whether follow-up manipulation of the weights is necessary (such as using an alternative weighting model or weight trimming).

For diagnostic purposes, it is often more convenient to examine the weights after they have been normalized so that their sum equals the sample size. i.e.: by dividing each weight in the sample by the average of the weights ($w_i^* = w_i / \bar{w}$). When weights are normalized to sum to the sample size, they have the appealing property of being more or less informative as they deviate from 1. A weight smaller than 1 for an observation indicates that the weighting procedure considers this observation less informative than the average observation. Conversely, a weight larger than 1 suggests that this observation is more informative, on average, than other observations.

For instance, if we have weights based on gender and find that males have weights smaller than 1 while females have weights larger than 1, we can infer that our sample has an over-representation of males and an under-representation of females - an imbalance that the weights attempt to rectify.

It is helpful to look at the distribution of the weights. Looking at the KDE plot can help detect multimodal distribution (which might indicate clusters of users of higher/lower representativeness of the population). It is also helpful to look at basic summary statistics, such as the main quartiles (25\%, 50\%, and 75\%) as well as the proportion of weights above and below certain values (e.g., over 2 and under 0.5, along with other similar quantities). This can help identify which proportions of the responses might be over/under weighted. Such insights could lead to followup changes to the final weighting model. For example, if we find out a handful of users have weights that are extremely large we might decide to look at the skewed features. We might find a need to bucket some classes in a covariate together, remove some features from the weighting model, use weight trimming, or some other post-processing manipulation to the weights.

% TODO: reference the e2e example table example.

\subsubsection{Diagnostics for the outcome}
\label{subsec:diagnostics_outcome}

The entire procedure of fitting weights and diagnostics is geared towards an impactful change in the outcome (or outcomes) of interest towards reducing the estimation bias. A common population parameter of interest is the mean. The statistics used to review it are the sample weighted mean, the variance of the weighted mean, as well as asymptotic confidence intervals.

The formula for the weighted mean, using a Horvitz–Thompson estimator \cite{horvitz1952generalization}, is simply:

\begin{equation}
\bar{y}_w = \frac{\sum_{i=1}^{n} w_i y_i }{\sum_{i=1}^{n} w_i}
\end{equation}

The variance of the weighted mean is based on the $\pi$-estimator for the ratio-mean:\cite{sarndal1992model} 

\begin{equation}
\widehat {V (\bar y_w)} 
 = \frac{1}{(\sum_{i=1}^n w_i)^2} \sum_{i=1}^n w_i^2 (y_i - \bar y_w)^2
\end{equation}

This estimator works for cases when the probability of selection for each $y_i$ are not identical, treating the $y_i$ values themselves as fixed.\footnote{The formula presented for the variance of the weighted mean assumes that the weights are known and fixed quantities. Hence, this formula does not account for the uncertainty that is introduced from the estimation of the weights. If measuring this uncertainty is of interest, then it is possible to perform an end to end bootstrap simulation which includes re-sampling from the sample, calculating the weighted mean estimation, and then repeating the process a few times, and using the bootstrap estimations of the mean to estimate the variance.} See section  \ref{subsec:var_of_weighted_mean} for more details.

The confidence intervals (CI) available uses the above formula and are the standard approximate CI based on the central limit theorem:

\begin{equation}
CI(\mu): \bar{y}_w \pm z_{\alpha/2} \sqrt{\widehat {V (\bar y_w)} }
\end{equation}

In an applied setting, it is advisable to calculate the weighted mean and their CI after applying the weights, and also without weights, and compare the quantities to each other. The difference of the weighted and unweighted mean could be thought of as an estimator of the potential bias reduced by using the weights (assuming the general trend of the ASMD calculations on the covariates indicate a positive improvement in their imbalance). This estimated bias can be compared to the effective sample size to allow a rough decision if the increase in variance due to the weights is adequately compensated by the reduction in bias.

% TODO: reference. "Examples (from simulated data) of weighted means and their confidence intervals (CI)}"

\section{The \texttt{balance} workflow}
\label{sec:balance_workflow}

% We start be describing the \texttt{balance} workflow. 

\subsection{The workflow}
\label{subsec:balance_workflow}

% Following the Total Survey Error framework, surveys responses are often biased due to coverage error, sampling error and non-response bias (see subsection \ref{subsec:total_survey_error}, \nameref{subsec:total_survey_error}, for more details). Weighting is often an important step when analyzing survey data. For each unit in the sample (e.g. respondent to a survey), we attach a weight that can be understood as the approximate number of people from the target population that this respondent represents.

Survey data weighting using \texttt{balance} is achieved with the following three main steps:

\begin{enumerate}
    \item \textbf{Understanding the initial bias in the data relative to a target population}: First, the survey data is loaded for both respondents and the target population. A pandas DataFrame can be created using \texttt{pandas.read\_csv()} and converted into a \texttt{balance} \texttt{Sample} class object with \texttt{Sample.from\_frame}. A similar step is repeated for the target population's data, and then the two \texttt{Sample} objects can be combined by assigning the target object as the target of the sample object. Once the data is loaded, we can conduct a diagnostic evaluation of the sample-vs-target covariates' distributions to determine if weighting is necessary. These include ASMD and distribution plots such as bar-charts and kernel-density-estimation plots.
    \item \textbf{Adjusting the sample to the target}: next, we generate weights for the sample to more accurately represent the target population's distributions. Currently, the package implements the following methods: Inverse Probability Weighting (IPW) using LASSO regression, Covariate Balancing Propensity Score (CBPS), Post-stratification, and raking. These are all available through the \texttt{adjust} method in the \texttt{Sample} class.
    \item \textbf{Results evaluation}: Once the weights are estimated, their effect is evaluated on the covariates imbalance (again, using ASMD and plots), the effective sample size, and the change in weighted mean of the outcome as well as their confidence intervals.
\end{enumerate}

The next section gives a detailed example for applying this workflow.

\subsection{An end-to-end example}

\subsubsection{Understanding the initial bias}

\noindent \textbf{Loading simulated data}

This section presents an example of simulated data extracted from the \texttt{balance} tutorial page \cite{balance_docs_quickstart_tutorial}. The data set is comprised of two pandas DataFrames: one for the target population and the other for the sample population. Both DataFrames contain an identifier column (id), three covariate columns (gender, age\_group, and income), and an outcome variable (happiness)\footnote{Code for creating the distributions is available here: \url{https://github.com/facebookresearch/balance/blob/main/balance/datasets/__init__.py\#L17}.}

In this particular simulation, we intentionally designed the outcome to be associated with all covariates, ensuring that this relationship remains consistent for both the target and sample populations. It is important to note that in real-world data sets, we generally don't observe the outcome for the target population. However, in this simulated data set we have included it for illustrative purposes. This setup allows us to later demonstrate how weighting methods can mitigate bias and approximate population-level parameters more accurately.

In real-world use-cases the data is often loaded using \texttt{pandas.read\_csv()}. Here, we use pre-made DataFrames that can be loaded (and inspected) using the following Python code:

\begin{lstlisting}[language=Python]
from balance import load_data
# INFO (2023-05-14 09:00:15,410) [__init__/<module> (line 52)]: Using balance version 0.9.0

target_df, sample_df = load_data()

print("sample_df: \n", sample_df.head())
\end{lstlisting}

\begin{verbatim}
sample_df: 
  id  gender age_group     income  happiness
0  0    Male     25-34   6.428659  26.043029
1  1  Female     18-24   9.940280  66.885485
2  2    Male     18-24   2.673623  37.091922
3  3     NaN     18-24  10.550308  49.394050
4  4     NaN     18-24   2.689994  72.304208

# The target_df DataFrame looks similarly to sample_df.
\end{verbatim}

\noindent \textbf{Creating instances of the Sample class with the DataFrames}

The main class for our analyses is the \texttt{Sample} class from the \texttt{balance} package. The following illustrates how we incorporate the DataFrames into this class:

\begin{lstlisting}[language=Python]
from balance import Sample

sample = Sample.from_frame(sample_df, outcome_columns=["happiness"])

target = Sample.from_frame(target_df, outcome_columns=["happiness"])
# Usually the code will be simply:
# target = Sample.from_frame(target_df)
# This is since most times we do not have the outcome for the target. In the example in this paper we have added it just to validate later that the weights indeed help us reduce the bias of the outcome.


# Following this, we associate the Sample object instance of sample with that of the target object, enabling us to adjust the sample to match the target.
sample_with_target = sample.set_target(target)
\end{lstlisting}

The \texttt{Sample} class provides a wide range of attributes, methods, and properties. For instance, the \texttt{df} property can reveal the DataFrame encapsulated within the instance of the \texttt{Sample} class (e.g.: \texttt{sample\_with\_target.df}):

Invoking the \texttt{Sample} object directly provides a concise summary of its attributes:

\begin{lstlisting}[language=Python]
sample_with_target
\end{lstlisting}

\begin{verbatim}
(balance.sample_class.Sample)

        balance Sample object with target set
        1000 observations x 3 variables: gender,age_group,income
        id_column: id, weight_column: weight,
        outcome_columns: happiness
        
            target:
                 
	        balance Sample object
	        10000 observations x 3 variables: gender,age_group,income
	        id_column: id, weight_column: weight,
	        outcome_columns: happiness
	        
            3 common variables: gender,age_group,income
\end{verbatim}

\noindent \textbf{Exploring the imbalances in covariates}

We can use methods such as \texttt{.covars()} with \texttt{.plot()}, \texttt{.mean()}, and \texttt{.asmd()} to get some diagnostics about the imbalance.

We can use the \texttt{.plot()} method to look at the distributions of covariates in the sample versus the target data.

\begin{lstlisting}[language=Python]
sample_with_target.covars().plot()
\end{lstlisting}

\begin{figure}[htbp]
\centering
  \begin{subfigure}[t]{0.85\textwidth}
    \includegraphics[width=\linewidth]{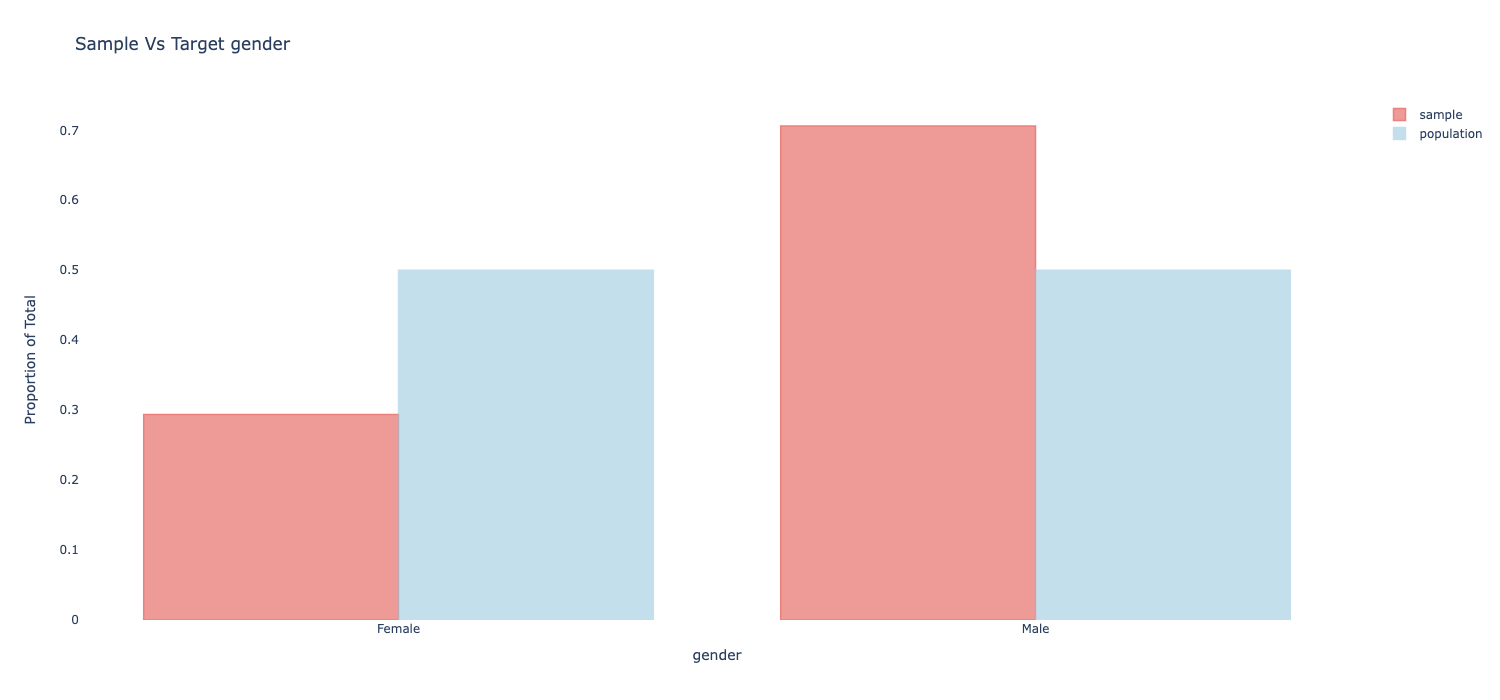}
    \caption{Bar plot for gender}
  \end{subfigure}
  
  \begin{subfigure}[t]{0.85\textwidth}
    \includegraphics[width=\linewidth]{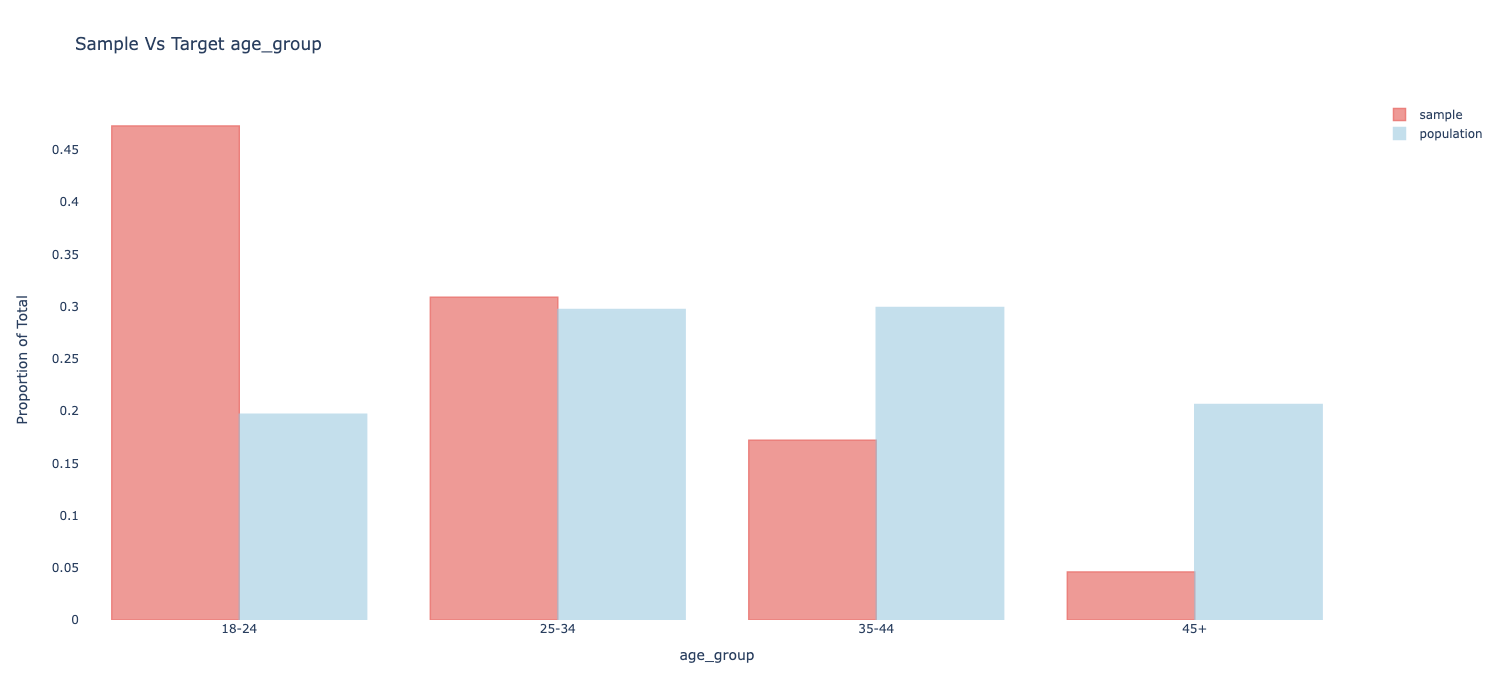}
    \caption{Bar plot for age group}
  \end{subfigure}
  
  \begin{subfigure}[t]{0.85\textwidth}
    \includegraphics[width=\linewidth]{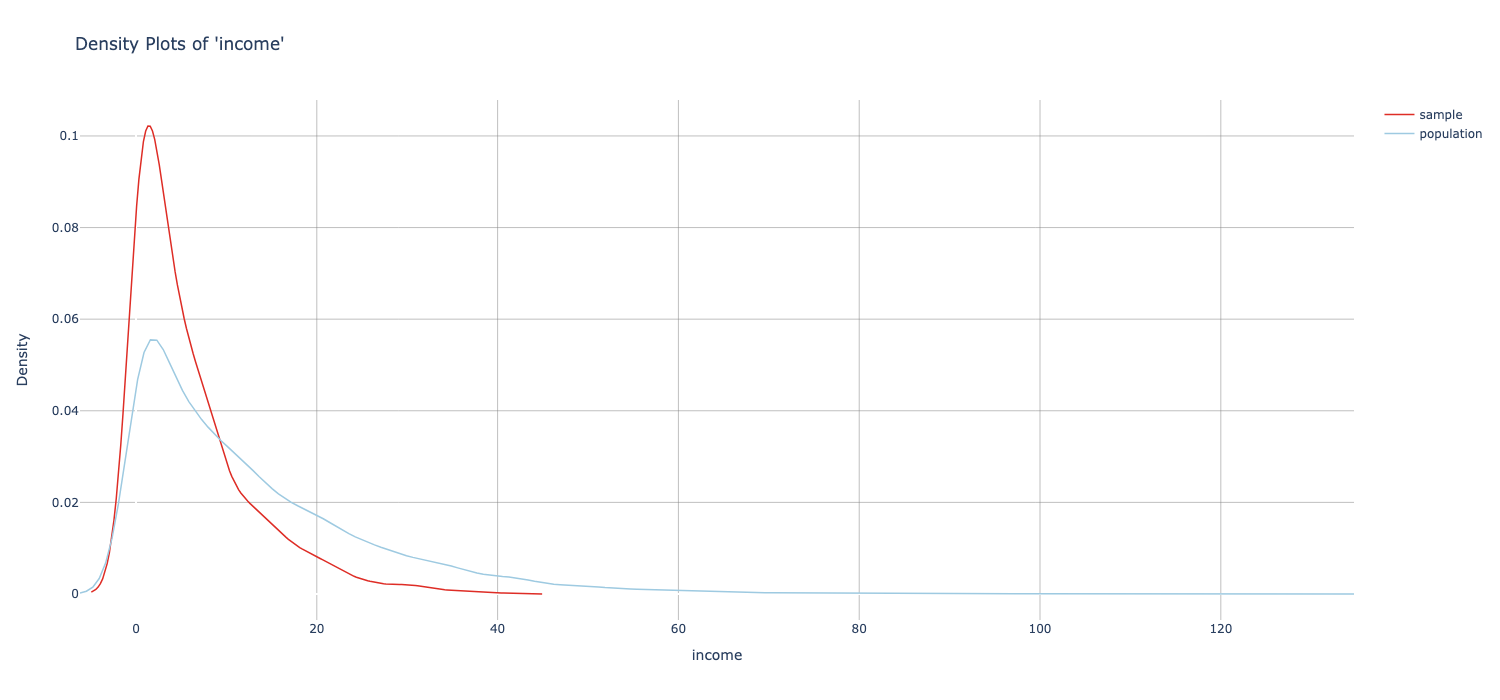}
    \caption{KDE for income}
  \end{subfigure}

\centering
\scriptsize % Adjust the font size here
\begin{tabular}{cccc}
    & \textcolor{col_unadjusted}{\rule[0.5ex]{1em}{1em}} & % \textcolor{col_adjusted}{\rule[0.5ex]{1em}{1em}} & 
    \textcolor{col_target}{\rule[0.5ex]{1em}{1em}} \\
    & Un-weighted sample & Weighted sample \\ % & True value \\
\end{tabular}
  
  \caption{Examples (from simulated data) of diagnostic plots for covariates (unweighted sample vs target)}
  \label{fig:sim_data_covar_plots_unweighted}
\end{figure}

The output in Figure \ref{fig:sim_data_covar_plots_unweighted} helps to easily identify imbalance. For example, we can see the sample has many more males than females, as opposed to a 50\%-50\% split in the target population. And for age\_group we can see how the sample is skewed towards younger respondents, as compared to the target population.

The package leverages plotly \cite{plotly} (as the default) to create interactive visualizations, but it also supports static figures using the seaborn package \cite{michael_waskom_2017_883859} for added flexibility.

The default \texttt{asmd} method uses ASMD to compare sample (which is unweighted) with the target using dummy variables for categorical variables, and calculates the ASMD for each of them. The aggregate ASMD per covariate can be achieved using the \texttt{aggregate\_by\_main\_covar = True} argument, as described in section \ref{subsubsec:diagnostics_covars_asmd}.

\begin{lstlisting}[language=Python]
print(sample_with_target.covars().asmd(aggregate_by_main_covar = True).T.round(2))
\end{lstlisting}

\begin{verbatim}
source         self
age_group      0.23
gender         0.25
income         0.49
mean(asmd)     0.33
\end{verbatim}

The ASMD helps quantify the levels of imbalance in each covariate.

\subsubsection{Fitting survey weights}

In order to estimate weights for the sample the \texttt{.adjust()} method as used on the \texttt{Sample} object. The default is ipw, and other methods could be invoked using the \texttt{method} argument.

\begin{lstlisting}[language=Python]
# Using ipw to fit survey weights
adjusted = sample_with_target.adjust()
\end{lstlisting}

\subsubsection{Evaluating the Results}

\noindent \textbf{Covariates}

We can get a basic summary of the results using the \texttt{.summary()} method:

\begin{lstlisting}[language=Python]
print(adjusted.summary())
\end{lstlisting}

\begin{verbatim}
Covar ASMD reduction: 59.7%, design effect: 1.897
Covar ASMD (7 variables): 0.327 -> 0.132
Model performance: Model proportion deviance explained: 0.172
\end{verbatim}

It shows that the weights led to an improvement of around 60\% reduction in the mean ASMD (from 0.327 to 0.132), and that the price we paid for it is an increasing the variance of the estimator by 1.897 in comparison to a random sample (as calculated using Kish's design effect, if assuming haphazard weights).

The same tools used to evaluate the bias before adjustment can be used for evaluating the effect of the weights on the balance after adjustment.

\begin{lstlisting}[language=Python]
adjusted.covars().plot()
\end{lstlisting}

\begin{figure}[htbp]
\centering
  \begin{subfigure}[t]{0.85\textwidth}
    \includegraphics[width=\linewidth]{fig_covar_plot_gender.png}
    \caption{Bar plot for gender}
  \end{subfigure}
  
  \begin{subfigure}[t]{0.85\textwidth}
    \includegraphics[width=\linewidth]{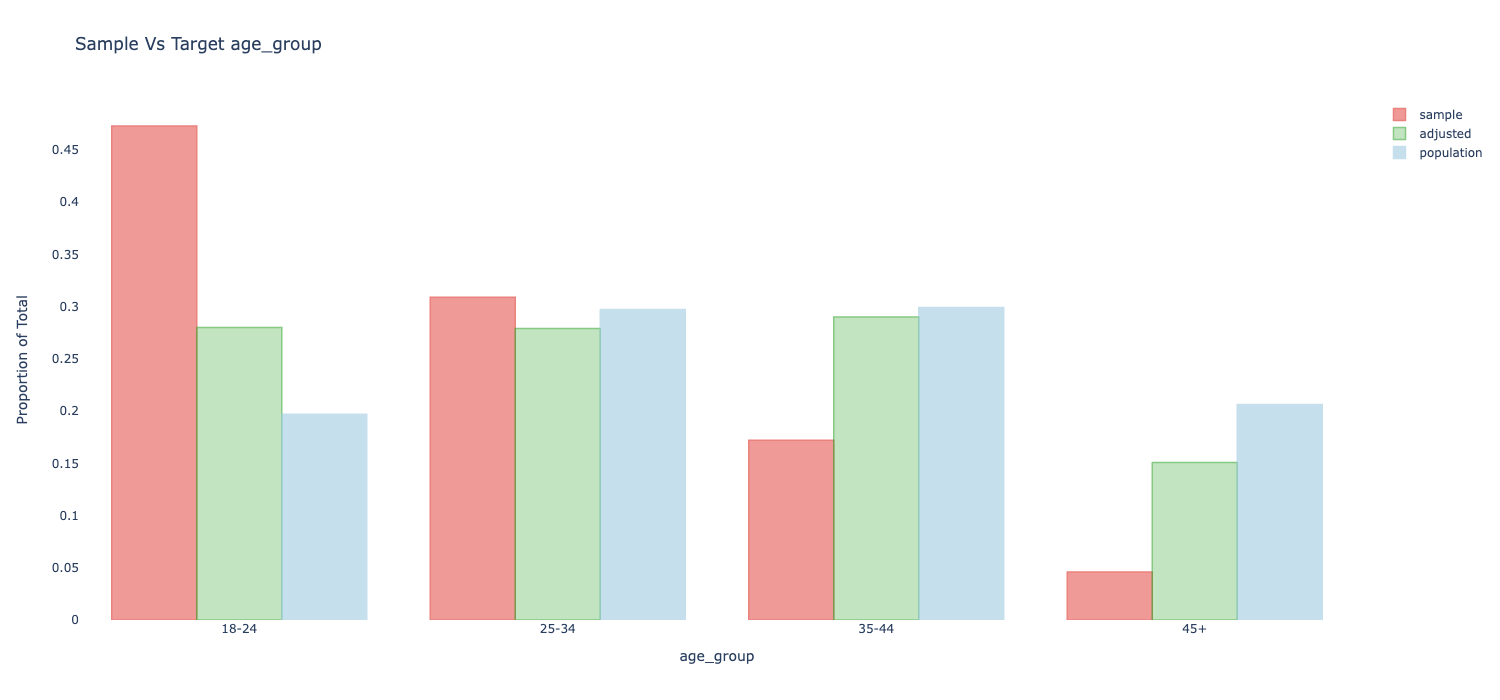}
    \caption{Bar plot for age group}
  \end{subfigure}
  
  \begin{subfigure}[t]{0.85\textwidth}
    \includegraphics[width=\linewidth]{fig_covar_plot_income.png}
    \caption{KDE for income}
  \end{subfigure}

\centering
\scriptsize % Adjust the font size here
\begin{tabular}{cccc}
    & \textcolor{col_unadjusted}{\rule[0.5ex]{1em}{1em}} & 
    \textcolor{col_adjusted}{\rule[0.5ex]{1em}{1em}} & 
    \textcolor{col_target}{\rule[0.5ex]{1em}{1em}} \\
    & Un-weighted sample & Weighted sample & True value \\
\end{tabular}
  
  \caption{Examples (from simulated data) of diagnostic plots for covariates (unweighted and weighted sample vs target)}
  \label{fig:sim_data_covar_plots}
\end{figure}

The output in Figure \ref{fig:sim_data_covar_plots} shows how the weights help mitigate some (though not all) of the bias, for all three covariates (gender, age and income).

We can also see the improvement per caovariate (averaged across category) using the \texttt{.asmd()} method:

\begin{lstlisting}[language=Python]
print(adjusted.covars().asmd(aggregate_by_main_covar = True).T.round(2))
\end{lstlisting}

\begin{verbatim}
source      self  unadjusted  unadjusted - self
age_group   0.06        0.23               0.18
gender      0.10        0.25               0.16
income      0.24        0.49               0.25
mean(asmd)  0.13        0.33               0.20
\end{verbatim}

We can see that while we got improvements in all covariates, there is still some imbalance that remained, especially in the income variable.

\noindent \textbf{Weights}
\label{subsubsubsection:weights}

Next, we wish look at the diagnostics of the weights to identify if there are any extreme weights or signs of issue that requires further investigation. This can be done by using the \texttt{summary} method on the \texttt{.weights()} method of the adjusted object.

\begin{lstlisting}[language=Python]
print(adjusted.weights().summary().round(2))
\end{lstlisting}

\begin{verbatim}
                                var       val
0                     design_effect      1.90
1       effective_sample_proportion      0.53
2             effective_sample_size    527.04
                                ...
7                      describe_min      0.31
11                     describe_max     11.65
16                      prop(w < 1)      0.65
21                    prop(w >= 10)      0.00
\end{verbatim}

We can see a design effect of 1.9 which corresponds with an effective sample size proportion of 53\%. Since the size of the sample was 1000, it means that the effective sample size is 527.
We can also see that 65\% of the weights are below 1, meaning that we down-sized 65\% of our sample. The minimal weight is 0.31 and the max weight is 11.65, with almost no weights above 10. A conclusion here is that the weights are not too extreme and we get some sense of the cost that using the weights would incur on the precision of our estimates.

\noindent \textbf{Outcome}

The \texttt{summary} method on the \texttt{outcomes} method gives us the weighted means and confidence intervals for the adjusted sample, the target, and the unadjusted sample data.

From the results below we can see that the real population level mean of happiness in the simulation was 56.2. In our (unweighted/unadjusted) sample it was 48.5. Meaning, the bias was roughly 7.7 points. After applying the weights, we got a value of 53.3, reducing the bias to roughly only 2.9 points. Note that this comparison is only possible in a simulated environment and is given here for a proof of concept of the effect of the weights. We can also see that the CI of the self and unadjusted show very different ranges of bands, indicating how the weights clearly got us a significant change in the estimated mean\footnote{Comparing the CI of the data with and without the weights is a good approximation for the impact of the weights, but is not statistically precise. Future work is planned for introducing more formal confidence intervals of the impact of the weights by using paired t-test style analysis. See the discussion and future work section for more.}. While the model improved the bias, we know it didn't fix it completely. This is because the model also did not perfectly fix the covariate imbalance, since it used some regularization in the process.

\begin{lstlisting}[language=Python]
print(adjusted.outcomes().summary())
adjusted.outcomes().plot()
\end{lstlisting}

\begin{table}[h]
\centering
\begin{tabular}{|c|c|}
\hline
Source & Happiness \\
\hline
Self & 53.389 \\
Target & 56.278 \\
Unadjusted & 48.559 \\
Self CI & (52.183, 54.595) \\
Target CI & (55.961, 56.595) \\
Unadjusted CI & (47.669, 49.449) \\
\hline
\end{tabular}
\end{table}

\begin{figure}[htbp]
\centering
    \includegraphics[width=\linewidth]{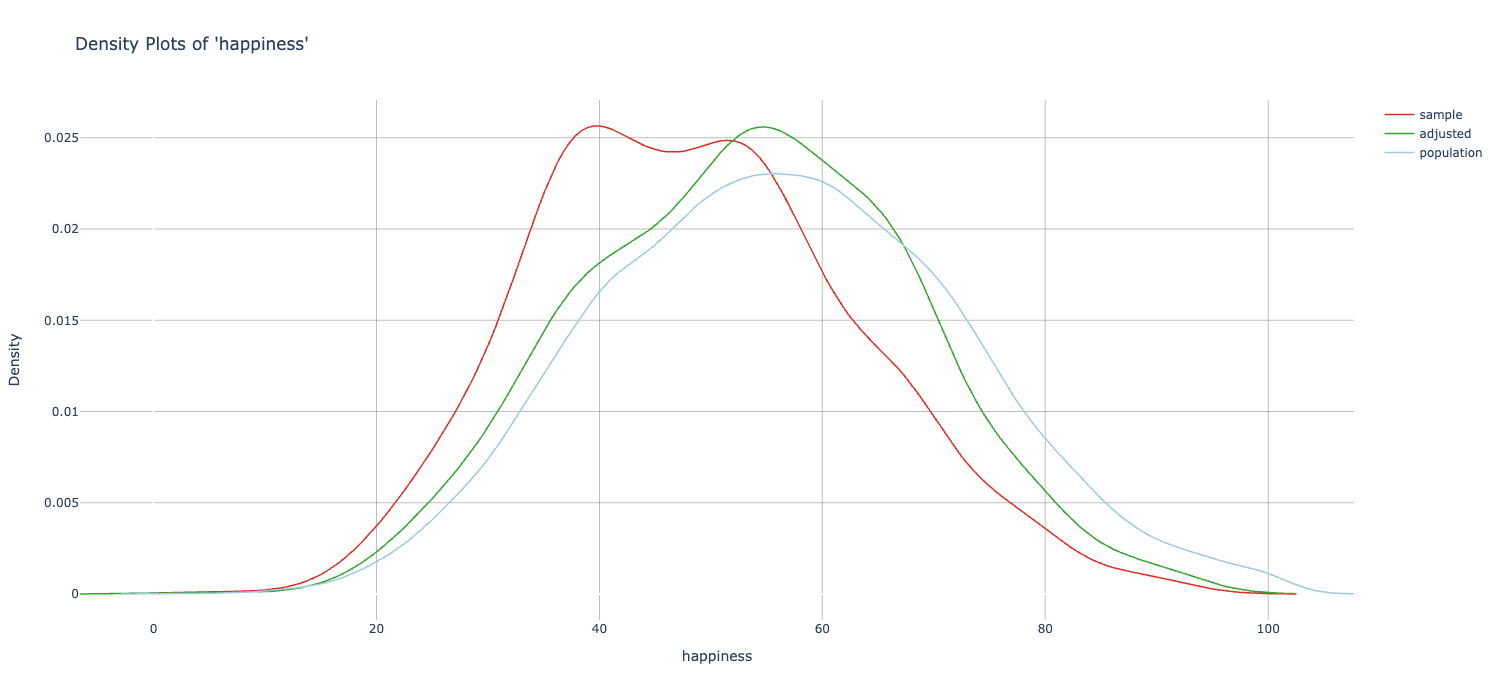}

\centering
\scriptsize % Adjust the font size here
\begin{tabular}{cccc}
    & \textcolor{col_unadjusted}{\rule[0.5ex]{1em}{1em}} & 
    \textcolor{col_adjusted}{\rule[0.5ex]{1em}{1em}} & 
    \textcolor{col_target}{\rule[0.5ex]{1em}{1em}} \\
    & Un-weighted sample & Weighted sample & True value \\
\end{tabular}
  
  \caption{Examples (from simulated data) of diagnostic plots for outcome (unweighted and weighted sample vs target)}
  \label{fig:sim_data_outcome_plots}
\end{figure}

The output of \texttt{.plot} is in Figure \ref{fig:sim_data_outcome_plots}. It shows that we got a relatively symmetrical uni-modal distribution (before and after applying the weights). So we don't observe and strong irregular behavior of the outcome. Note that we are able to compare the outcome with and without the weights in the sample to the real outcome distribution in the target population only because this is simulated data. In real-world cases, we are not expected to have access to the outcome distribution of the target population. Also, it is relatively common to get outcome responses in binary or likert scales, and not a continuous variable. The \texttt{.plot} would work with these just as well.

\noindent \textbf{Downloading data}

Once we are settled with the weights we got, we can download them as csv, as follows:

\begin{lstlisting}[language=Python]
adjusted.to_download()  # Will create a download link in jupyter
# We can also prepare the data to be exported as csv
# The following code showes the first 500 characters for simplicity:
adjusted.to_csv()
\end{lstlisting}

% \subsection{Technical details}
\subsection{How does \texttt{balance} implement the adjustment?}
\label{subsec:How does balance implement the adjustment?}

\textbf{Pre-processing}
Before applying any of the adjustment methods, \texttt{balance} performs a pre-processing step to improve models' results. The pre-processing step includes a few best practiced that makes the use of \texttt{balance} easy and automatic for a default usage. 

\noindent \textbf{Transformations.} \texttt{balance} applies the following default behaviours:

\begin{enumerate}

    \item Handling missing values: \texttt{balance} handles missing values automatically by adding a special indicator column to any variable that contains missing values. The advantage of this is that these are then considered as a separate category for the adjustment. 
    
    \item Feature engineering: by default, \texttt{balance} applies feature engineering to be able to fit the covariate distribution better, and not only the first moment. Specifically, each continuous variable is bucketed into 10 quantiles buckets, and rare categories variables are grouped together so to avoid overfitting\footnote{The user has also an option to change these default behaviours, through setting different values to the \texttt{transformations} argument of \texttt{Sample.adjust}.}.

\end{enumerate}

\noindent \textbf{Model matrix.} The model matrix of the covariates used in \texttt{balance} for the logistic regression in ipw and for CBPS propensity score is constructed before the fitting is done using the transformed variables and one-hot encoding for discrete variables. The default behaviour is an additive model including all joint covariates of the target and the observed sample. However, thorough the argument \texttt{formula}, one can input a formula for a specified relation between the variables. The formulas adopt the notation from the \texttt{patsy} Python package \cite{patsy}, facilitating a range of operations like addition, multiplication (for interaction effects), and power transformations. A detailed example is available in the \href{https://import-balance.org/docs/tutorials/balance_transformations_and_formulas/}{"balance: transformations and formulas"} tutorial \cite{balance_transformations_and_formulas_tutorial}.

\noindent \textbf{Adjustment through \texttt{ipw}}

\texttt{ipw} is implemented using LASSO regularized logistic regression. 
To avoid non-balanced categories in the logistic regression, \texttt{balance} scales the prevalence of the target population to be similar to the observed sample.

The penalty factor $\lambda$ of the LASSO is chosen through cross-validation. Two methods for choosing the parameter are suggested:
\begin{enumerate}

    \item Unbounded Design Effect: If one doesn't want to bound the design effect of the resulted weights (the default behaviour with \texttt{max\_de=None}), the penalty if chosen using \texttt{lambda\_1se}, which is the largest value of $\lambda$ such that the cross-validated error is within one standard error of the minimum value.
    
    \item Bounded Design Effect: If one chooses to bound the design effect (e.g. by using \texttt{max\_de=2}), a grid search over 10 of the values of $\lambda$ that brings the largest design effect within the bound is done, where the  $\lambda$ is chosen to be the one that brings the largest ASMD reduction. 
\end{enumerate}

In addition, a \texttt{penalty\_factor} argument can be also used to indicate how much the model should focus to adjust each term of the formula. Larger penalty factors means that the covariate is more likely to be regularized by the LASSO penalty and as a result the adjustment of this covariate will be smaller, i.e. will end in a less balanced covariate. This feature can be particularly useful when certain components are believed to be more or less responsible for bias in the data, or when the user wants to explore different adjustment scenarios.

\noindent \textbf{Post-processing}

Weights in \texttt{balance} are scaled to the population size after estimated, and can be interpret as the number of people from the target the sample unit represent. After the adjustment and scaling is done, weights trimming from above is performed. This is done in order avoid over fitting of the model and unnecessary variance inflation. The weights are trimmed and scales in a way that keeps the mean and sum of the weights the same as before trimming, such that the interpretation how many units in the target this unit represent holds after trimming.

\section{Future directions}
\label{sec:future_direction}

The \texttt{balance} package offers benefits for researchers interested in analyzing data with non-response bias in the Python environment by being easy to use, providing an end-to-end workflow, and released as open-source. While comprehensive, there is still room for improvement and expansion. This section highlights several possible areas for future development in the \texttt{balance} package.

\begin{enumerate}
    % \item \textbf{Transition from glmnet to \texttt{sklearn}:} Moving to \texttt{sklearn} could provide several benefits, such as support for Windows OS, the ability to use more models beyond GLM LASSO (like random forest), the ability to move from GPL to MIT licence, and compatibility with Python 3.11 and later versions.
    \item \textbf{Better Diagnostic Tools for Covariates:} The current metric of ASMD has limitations, especially when applied to a wide range of distributions and for categorical variables. Future versions could include more robust measures like the Common Language Effect Size \cite{mcgraw1992common} and better methods for handling categorical variables, such as Kullback-Leibler divergence. There is also room for adding statistical hypothesis tests for the evaluations, as well as more plots. The \texttt{cobalt} R package \cite{greifer2020covariate} is a good source of inspiration.
    \item \textbf{Expanded Estimation and Diagnostic Tools for Outcomes:} Currently, the package primarily provides the weighted mean and its confidence intervals. A helpful improvement would be to directly measure the estimated bias reduction caused by the weights, including a confidence interval for this estimate. Also, current implementation focuses on the weighted average and the linearization (Taylor) estimator for the variance. Other possible statistics, and estimations of variance exists. The \texttt{samplics} package already implements some of these and would be a good source of inspiration \cite{diallo2021samplics}.
    \item \textbf{Diagnostics for the bias-variance trade-offs when using weights:} At present, the user is provided with a set of weights but with no easy way to check the bias-variance tradeoffs for alternative levels of trimming or tuning other parameters. A future version of the package could include more diagnostics tools and allow automated functions for weight trimming, such as based on empirical-MSE estimation for a given outcome over a range of potential weight trimming values. This could lead to a better balance between the variance induced by the weights and the bias they reduce and save researcher's time in manual tweaking.
    \item \textbf{Built-in Model Comparison for Multiple Weights:} Our ultimate goal is to allow the most flexibility to the user by conducting easy comparisons of multiple models and adjustments to the weights in order to choose the model that best fits his/hers data.
    \item \textbf{Feature Selection for Propensity Score Models:} When given several potential models, it can be challenging to choose the best one. This choice could depend on various factors, such as the balance between reduced bias and incurred variance or the impact of different models on different outcomes. Further development in this area could provide useful tools for sensitivity analysis and decision making.
    \item \textbf{Expansion Beyond Propensity Score Models:} The next step for the package could be to include outcome models and doubly robust models. Thus making the package more versatile and comprehensive.
\end{enumerate}

These possible improvements represent exciting opportunities for the future of the \texttt{balance} package, aiming to provide a more robust and user-friendly tool for researchers in the Python environment. We welcome any feedback, suggestions, and opportunities for collaborations.

\clearpage % Start a new page for the References section
\phantomsection % Create a phantom section for accurate TOC page number
\addcontentsline{toc}{section}{References} % Add "References" to TOC

\bibliographystyle{ieeetr} % We choose the "ieeetr" reference style
% https://www.overleaf.com/learn/latex/Bibtex_bibliography_styles
\bibliography{refs} % Entries are in the refs.bib file

\newpage
\appendix
\appendixpage

% \addappheadtotoc
%\addtocontents{toc}{\protect\setcounter{tocdepth}{0}}

\section{Acknowledgments}

The \texttt{balance} package was (and is) developed by many people, including: Roee Eilat, Tal Galili, Daniel Haimovich, Kevin Liou, Steve Mandala, Adam Obeng (author of the initial internal Meta version), Tal Sarig, Luke Sonnet, Sean Taylor, Barak Yair Reif, and others.

The \texttt{balance} package was open-sourced by Tal Sarig, Tal Galili and Steve Mandala, from Central Applied Science at Meta, in late 2022.

Branding created by Dana Beaty, from the Meta AI Design and Marketing Team.

\section{Limitations of the ASMD}
\label{subsec:limitations_of_asmd}

It is also worth noting some of the disadvantages in ASMD:

\begin{enumerate}
\item Sensitivity to extreme values: Since ASMD is based on the first moment, ASMD can be sensitive to outliers or extreme values in the data, which can lead to a distorted representation of the differences between the two groups. This could be mitigated by turning to robust measure, but these are currently not implemented in \texttt{balance}.
\item Inability to detect distributional differences: ASMD focuses solely on the mean difference between the two groups, and does not account for differences in other distributional characteristics, such as variance, skewness or number of modes. This means that two groups with similar means but different variances or shapes may have a low ASMD value, which could be misleading. This can be addressed by looking at distribution plots. The next section discusses methods that are available in \texttt{balance}.
\item The need for context: ASMD values are unitless and can be difficult to interpret without context. Does an ASMD value below 0.1 indicate an effect size which is small or large? The interpretation of ASMD is often comparative within a specific research context. For example, ASMD changes could be compared across different covariates and alternative weights, so to identify which set of weight effects the bias of which covariate.
\item Limited applicability to categorical variables: ASMD is primarily designed for comparing continuous variables, and its applicability to categorical variables is more limited. In such cases the covariate can be turned into several dummy variables using one hot encoding and the ASMD can be calculated on these values of zeros and ones. Alternative measures that directly compare categorical distributions \cite{math11091982} are currently not implemented in \texttt{balance}. 
\end{enumerate}

Despite these limitations, the ASMD can be a useful measure for comparing the effect of the weights on the covariates.

\section{Kish's design effect}
\label{subsec:kishs_deff_appendix}

\appendixsubsection{Design effect in general}

A \emph{design effect} \cite{wiki:design_effect}\footnote{The text in this section is a modified version of the text we wrote for the Wikipedia article on Kish's design effect \cite{wiki:design_effect}.} is a measure of the increase in variance of an estimate due to the use of survey weights compared to an equal probability sample of the same size. Theoretically, it is calculated as follows:

\begin{equation}
D_{eff} = \frac{Var(\hat \theta_{weighted})}{Var(\hat \theta_{un-weighted})}
\end{equation}

where $Var_{weighted}$ and $Var_{unweighted}$ are the variances of the weighted and unweighted estimates, respectively.

A design effect greater than 1 indicates that the variance of the weighted estimate is larger than that of an unweighted estimate, while a design effect less than 1 suggests that the variance of the weighted estimate is smaller (for example, when using design weights based on stratified sampling). A design effect of 1 implies that the use of weights does not affect the variance of the estimate (which happens only if all weights are equal to the same, non 0, value).

A design effect that is larger than 1 does not necessarily imply that the weights are undesirable, as they can still improve the accuracy and representativeness of the estimates. Put differently, it may be that the bias corrected by applying the weights is substantially larger than the variance added due to using them.

Kish's design effect is a specific measure for a specific parameter (the population mean), with specific assumptions. The following sections discusses these some of these assumptions.

\appendixsubsection{Assumptions and derivation}

The formula of Kish's design effect computes the increase in variance of the weighted mean due to "haphazard" weights, which occur when y consists of observations selected using unequal probabilities, without within-cluster correlation or any relationship to the expected value or variance of the outcome measurement. From a model-based perspective \cite{gabler1999model}, the formula holds when all $n$ observations ($y_{1},...,y_{n}$) are (at least approximately) uncorrelated and have the same variance for some response variable of interest ($y$). The formula also assumes that the weights are not random variables but rather known constants. These, for example, can be the inverse of the selection probability for a pre-determined and known sampling design.

The conditions on $y$ are trivially satisfied if the $y$ observations are independent and identically distributed (i.i.d.) with the same expectation and variance. It is important to note that if $y_{1},...,y_{n}$ do not have the same expectations, the estimated variance of the estimator cannot be used for calculating the variance using a simple weighted variance formula, as the estimation assumes that all $y_{i}$s have the same expectation. Specifically, if there is a correlation between the weights and the outcome variable $y$, the expectation of $y$ is not the same for all observations but rather depends on the specific weight value for each observation. In such cases, while the design effect formula might still be accurate (assuming other conditions are met), a different estimator for the variance of the weighted mean may be needed, such as a weighted variance estimator.

If different $y_{i}$'s have distinct variances, the weighted variance might capture the correct population-level variance, but Kish's formula for the design effect may no longer be valid. A similar issue occurs if there is a correlation structure in the samples, such as when using cluster sampling.

Kish's formula estimates the increase in the variance of the weighted mean based on "haphazard" weights. Let $y$ be observations selected using unequal selection probabilities (with no within-cluster correlation, and no relationship to the expectancy or variance of the outcome measurement);\cite{kish1992} and let $y'$ be the observations we would have had if we got them from simple random sample, then Kish's formula for Deff is:

$D_{eff (kish)} =
\frac{var\left(\bar{y}_w\right)}{var\left(\bar{y}'\right)} = 
\frac{ var\left(\nicefrac{ \left(\sum\limits_{i=1}^n w_i y_i\right)}{\left(\sum\limits_{i=1}^n w_i \right)} \right) }{ var\left( \nicefrac{\sum\limits_{i=1}^n y_i'}{n} \right)}$

From a model based perspective\cite{gabler1999model}, this formula holds when all $n$ observations ($y_1, ..., y_n$) are (at least approximately) uncorrelated ($\forall (i \neq j): cor(y_i, y_j) = 0$), with the same variance ($\sigma^2$) in the response variable of interest ($y$). It also assumes the weights themselves are not a random variable but rather some known constants (E.g.: the inverse of probability of selection, for some pre-determined and known sampling design).

The conditions on $y$ are trivially held if the y observations are i.i.d with the same expectation and variance. In such case we have $y=y'$, and we can estimate $var\left(\bar{y}_w\right)$ by using $\overline{var\left(\bar{y}_w\right)} = \overline{var\left(\bar{y}\right)} \times D_{eff}$ \cite{kish1992}.

\appendixsubsection{Proof}

We present here a simplified proof to Kish's formula: $D_{eff}:=\frac{var\left(\bar{y}_w\right)}{var\left(\bar{y}'\right)} =  \frac{\overline{w^2}}{ \bar{w}^2 } $
for the case when there are no clusters (i.e.: no intraclass correlation between the elements of the sample), so that each strata includes only one observation. The proof is shown in full in \cite{gabler1999model} .

\begin{equation}
\begin{split}
var\left(\bar{y}_w\right) 
& \stackrel{1}{=} var\left(\frac{ \sum\limits_{i=1}^n w_i y_i}{\sum\limits_{i=1}^n w_i} \right)
  \stackrel{2}{=} var\left( \sum\limits_{i=1}^n w_i' y_i \right)
  \stackrel{3}{=} \sum\limits_{i=1}^n var\left( w_i' y_i \right) \\ 
& \stackrel{4}{=} \sum\limits_{i=1}^n w_i'^2 var\left( y_i \right)
  \stackrel{5}{=} \sum\limits_{i=1}^n w_i'^2 \sigma^2
  \stackrel{6}{=} \sigma^2 \sum\limits_{i=1}^n w_i'^2 
  \stackrel{7}{=} \sigma^2 \frac{\sum\limits_{i=1}^n w_i^2}{\left( \sum\limits_{i=1}^n w_i\right) ^2}  \\
& \stackrel{8}{=} \sigma^2 \frac{\sum\limits_{i=1}^n w_i^2}{\left( \sum\limits_{i=1}^n w_i \frac{n}{n} \right) ^2 } 
  \stackrel{9}{=} \sigma^2 \frac{\sum\limits_{i=1}^n w_i^2}{\left( \frac{\sum\limits_{i=1}^n w_i}{n} \right) ^2 n^2} 
  \stackrel{10}{=} \frac{\sigma^2}{n} \frac{\frac{\sum\limits_{i=1}^n w_i^2}{n}}{ \left( \frac{\sum\limits_{i=1}^n w_i}{n} \right) ^2 } \\
& \stackrel{11}{=} \frac{\sigma^2}{n} \frac{\overline{w^2}}{ \bar{w}^2 } 
  \stackrel{12}{=} var\left(\bar{y}'\right) D_{eff} \\
& \implies D_{eff (kish)} =\frac{var\left(\bar{y}_w\right)}{var\left(\bar{y}'\right)} 
\end{split}
\end{equation}

Transitions:

\begin{enumerate}
\item from definition of the weighted mean.
\item using normalized (convex) weights definition (weights that sum to 1): $w_i' = \frac{w_i}{\sum\limits_{i=1}^n w_i}$.
\item sum of uncorrelated random variables.
\item If the weights are constants (from the basic properties of the variance). Another way to say it is that the weights are known upfront for each observation i. Namely that we are actually calculating $var\left(\bar{y}_w | w \right) $
\item assume all observations have the same variance ($\sigma^2$).
\end{enumerate}

\section{Estimating the variance of the weighted mean}
\label{subsec:var_of_weighted_mean}

\appendixsubsection{Formulation}

This section discusses the derivation of the formula presented in the paper for the variance of the weighted mean, also known as $\pi$-estimator for ratio-mean.\footnote{The text in this section is a modified version of the text we wrote for the Wikipedia article on the weighted mean \cite{wiki:weighted_arithmetic_mean}.}

We are interested in estimating the variance of the weighted mean when the various $y_i$ are not assumed to be i.i.d random variables. An alternative perspective for this problem is that of some arbitrary sampling design of the data in which units are selected with unequal probabilities (with replacement) \cite{cochran1977sampling}.

Unlike classical ”model based” approaches, in which the randomness is described by the randomness of the $y$ value, here we consider the value of $y_i$ as constant, where the variability comes from the selection procedure. We let $R_i$ be the Bernoulli indicator that is equal to $1$ if observation $i$ is in the observed sample, and $0$ if not. The probability of a unit to be sampled given a sample $\mathcal{S}$ of size $n$ is denoted by $\pi_i:=P(R_i=1 \mid \mathcal{S})$. Furthermore, we denote the  one-draw probability of selection by $p_i := P(R_i=1 | \verb|one sample draw|) \approx \frac{\pi_i}{n}$. For the following derivation we'll assume that the probability of selecting each element is fully represented by these probabilities \cite{sarndal1992model}, i.e. selecting some element will not influence the probability of drawing another element (this doesn't apply for things such as cluster sampling design).
 
Since each outcome $y_i$ is fixed, and the randomness comes from unit $i$ being included in the sample or not ($R_i$), we often talk about the multiplication of the two, which is a random variable. To avoid confusion in what to follow, we define: $y'_i = y_i \cdot R_i$. This satisfies: $\mathbb{E}[y'_i] = y_i \mathbb{E}[R_i] = y_i \pi_i$ and $\mathbb{V}[y'_i] = y_i^2 \mathbb{V}[R_i] = y_i^2 \pi_i(1-\pi_i)$.

In this "design based" perspective, the weights are obtained by taking the inverse of the selection probability (i.e.: the inflation factor), i.e. $w_i = \frac{1}{\pi_i} \approx \frac{1}{n \times p_i}$. The weights in this setting are considered fixed and known.

We assume that the target population size $N$ is unknown, and is estimated by $\hat{N} = \sum_{i=1}^n w_i$. Our parameter of interest is the weighted mean, that can be written as a ratio: 
\begin{equation}
    \bar Y = \frac{\sum_{i=1}^N \frac{y_i}{\pi_i}}{\sum_{i=1}^N \frac{1}{\pi_i}} = \frac{\sum_{i=1}^N w_i y_i}{\sum_{i=1}^N w_i}
\end{equation} 
This ratio is estimated by the observed sample using:  

\begin{equation}
   \hat {\bar Y} = \frac{\sum_{i=1}^n \frac{y_i}{\pi_i}}{\sum_{i=1}^n \frac{1}{\pi_i}} = \frac{\sum_{i=1}^n w_i y'_i}{\sum_{i=1}^n w_i R_i}
\end{equation}

This is called a Ratio estimator and it is approximately unbiased for $\bar{Y}$\cite[p.182]{sarndal1992model}

In this case, the variability of the ratio depends on the variability of the random variables both in the numerator and the denominator - as well as their correlation. Since there is no closed analytical form to compute this variance, various methods are used for approximate estimation, primarily Taylor series first-order linearization, asymptotics, and bootstrap/jackknife.\cite[p. 172]{sarndal1992model} The Taylor linearization method could lead to under-estimation of the variance for small sample sizes in general, but that depends on the complexity of the statistic. For the weighted mean, the approximate variance is supposed to be relatively accurate even for medium sample sizes \cite{sarndal1992model}. For when the sampling has a random sample size, such as in Poisson sampling, it is as follows: \cite{sarndal1992model}

\begin{equation}
\widehat {V (\bar y_w)} 
 = \frac{1}{(\sum_{i=1}^n w_i)^2} \sum_{i=1}^n w_i^2 (y_i - \bar y_w)^2
\end{equation}

We note that if $\pi_i \approx p_i n$, then either using $w_i = \frac{1}{\pi_i}$ or $w_i = \frac{1}{p_i}$ would give the same estimator, since multiplying $w_i$ by some factor would lead to the same estimator. It also means that if we scale the sum of weights to be equal to a known-from-before population size $N$, the variance calculation would look the same. When all weights are equal to one another, this formula is reduced to the standard unbiased variance estimator.

Note that for the trivial case in which all the weights are equal to 1, the above formula is just like the maximum-likelihood formula for the variance of the mean (but not that it is not the unbiased variance, i.e. dividing it by n instead of (n-1)).

\appendixsubsection{Proof}
We show here a short proof for the variance formula presented above: 
\begin{equation}
\widehat {V (\bar y_w)} 
 = \frac{1}{(\sum_{i=1}^n w_i)^2} \sum_{i=1}^n w_i^2 (y_i - \bar y_w)^2
\end{equation}

The Taylor linearization states that for a general ratio estimator, $Q$, of two sums, $Y$ and $Z$, can be expressed by: \cite[p.178]{sarndal1992model}

\begin{equation}
\hat Q = \frac{\hat{Y}}{\hat{Z}} = \frac{\sum_{i=1}^n w_i y'_i}{\sum_{i=1}^n w_i z'_i} \approx Q + \frac{1}{Z} \sum_{i=1}^n \left( \frac{y'_i}{\pi_i} - Q \frac{z'_i}{\pi_i} \right)
\end{equation}

And the variance can be approximated by: \cite[p.178]{sarndal1992model}

\begin{equation}
\widehat {V (\hat Q)} =  \frac{1}{\hat{Z}^2} \sum_{i=1}^n \sum_{j=1}^n \left( \check{\Delta}_{ij} \frac{y_i - \hat Q z_i}{\pi_i}\frac{y_j - \hat Q z_j}{\pi_j} \right) = \frac{1}{\hat{Z}^2} \left[  \widehat {V (\hat Y)} + \hat Q \widehat {V (\hat Z)} -2 \hat Q \hat C (\hat Y, \hat Z) \right]
\end{equation}

where $ \hat C (\hat Y, \hat Z) $ is the estimated covariance between the $Y$ and $Z$, and $\Delta_{ij} = C(R_i, R_j)$.

Since $\hat C$ is the covariance of two sums of random variables, it would include many combinations of covariances that will depend on the indicator variables. If the selection probability are uncorrelated (i.e.: $\forall i \neq j: \Delta_{ij} = C(R_i, R_j) = 0$), this term would include only the summation of $n$ covariances for each element $i$ between $ y'_i = R_i \cdot y_i $ and $ z'_i = R_i \cdot z_i $. This helps illustrate that this formula incorporates the effect of correlation between $y$ and $z$ on the variance of the ratio estimators.

When defining $ z_i = 1 $ the above becomes: \cite[p.182]{sarndal1992model}

\begin{equation}
\widehat{V (\hat Q)} = \widehat {V (\bar y_w) }
 = \frac{1}{\hat{N}^2} \sum_{i=1}^n \sum_{j=1}^n \left( \check{\Delta}_{ij} \frac{y_i - \bar y_w}{\pi_i}\frac{y_j - \bar y_w}{\pi_j} \right)
\end{equation}

where $\check{\Delta}_{ij} = \frac{\Delta_{ij}}{\pi_{ij}}$. If the selection probability are uncorrelated (i.e.: $\forall i \neq j: \Delta_{ij} = C(R_i, R_j) = 0$), and when assuming the probability of each element is very small (i.e.: $(1- \pi_i) \approx 1$), then the above reduced to the following:

\begin{equation}
\widehat{ V (\bar y_w) }
 = \frac{1}{\hat{N}^2} \sum_{i=1}^n \left( (1- \pi_i) \frac{y_i - \bar y_w}{\pi_i} \right)^2
 = \frac{1}{(\sum_{i=1}^n w_i)^2} \sum_{i=1}^n w_i^2 (y_i - \bar y_w)^2.
\end{equation}

% \printbibliography

% \bibliographystyle{IEEEtran}
% \bibliography{IEEEabrv,mybibfile}

\end{document}